\documentclass[prd,superscriptaddress,
preprint,nofootinbib%
,amssymb, amsmath,mathrsfs,nobibnotes,aps,11pt]{revtex4-1}

\usepackage{epsfig}
\usepackage{graphicx}
 \usepackage{placeins}
\usepackage{mathrsfs}
\usepackage{color}
\usepackage{amssymb}
\usepackage{float}
\usepackage{verbatim}
\usepackage{amsmath}
\usepackage[toc,page]{appendix}
\usepackage[colorlinks=true,linkcolor=blue]{hyperref}
\expandafter\ifx\csname package@font\endcsname\relax\else
 \expandafter\expandafter
 \expandafter\usepackage
 \expandafter\expandafter
 \expandafter{\csname package@font\endcsname}%
\fi

\begin{document}
\title{Second order relativistic viscous hydrodynamics within an effective description of hot QCD medium}
\author{Samapan Bhadury}
\email{samapan.bhadury@niser.ac.in}
\affiliation{School of Physical Sciences, National Institute of Science Education and Research, HBNI, Jatni-752050, India.}
\author{Manu Kurian}
\email{manu.kurian@iitgn.ac.in}
\affiliation{Indian Institute of Technology Gandhinagar, Gandhinagar-382355, Gujarat, India.}
\author{Vinod Chandra}
\email{vchandra@iitgn.ac.in}
\affiliation{Indian Institute of Technology Gandhinagar, Gandhinagar-382355, Gujarat, India.}
\author{Amaresh Jaiswal}
\email{a.jaiswal@niser.ac.in}
\affiliation{School of Physical Sciences, National Institute of Science Education and Research, HBNI, Jatni-752050, India.}
%
%%%%%%%%%%%%%%%%%%%%%%%%%%%%%%%%%%%%%%%%%%%%%%%%%%%%%%%%%%%%%%%
\begin{abstract}
The second-order hydrodynamic equations for evolution of shear and bulk viscous pressure have been derived within the framework of covariant kinetic theory based on the effective fugacity quasiparticle model. The temperature-dependent fugacity parameter in the equilibrium distribution function leads to a mean field term in the Boltzmann equation which affects the interactions in the hot QCD matter. The viscous corrections to distribution function, up to second-order in gradient expansion, have been obtained by employing a Chapman-Enskog like iterative solution of the effective Boltzmann equation within the relaxation time approximation. The effect of mean field contributions to transport coefficients as well as entropy current has been studied up to second-order in gradients. In contrast to the previous calculations, we find non-vanishing entropy flux at second order. The effective description of relativistic second-order viscous hydrodynamics, for a system of interacting quarks and gluons, has been quantitatively analyzed in the case of the $1+1-$dimensional boost invariant longitudinal expansion. We study the proper time evolution of temperature, pressure anisotropy, and viscous corrections to entropy density for this simplified expansion. The second order evolution of quark-gluon plasma is seen to be affected significantly with the inclusion of mean field contributions and the realistic equation of state. 
\end{abstract}
\newpage
%%%%%%%%%%%%%%%%%%%%%%%%%%%%%%%%%%%%%%%%%%%%%%%%%%%%%%%%%%%%

\maketitle

%%%%%%%%%%%%%%%%%%%%%%%%%%%%%%%%%%%%%%%%%%%%%%%%%%%%%%%%%%%%
% 
\section{Introduction}
The heavy-ion collision experiments in Large Hadron Collider (LHC) and Relativistic Heavy Ion Collider (RHIC) have conclusively established the existence of locally thermalized hot nuclear matter called quark-gluon plasma (QGP), under extreme conditions of temperature/energy density~\cite{Adams:2005dq, Adcox:2004mh, Back:2004je, Arsene:2004fa, Aamodt:2010pb, Heinz:2008tv}. The phenomenological analysis of experimental observations suggests that the QGP formed in these collisions exhibit collective behaviour and can be considered as a strongly-coupled fluid with the smallest viscosity to entropy ratio~\cite{Kovtun:2004de, Luzum:2008cw}. The QGP is also inferred to be the most vortical fluid ever observed \cite{STAR:2017ckg}. Relativistic hydrodynamics has proved to be an efficient approach towards the description of the space-time evolution of the QGP and hence the observed collective phenomena in the heavy-ion collisions~\cite{Jeon:2015dfa, Florkowski:2017olj, Heinz:2013th, Braun-Munzinger:2015hba, Jaiswal:2016hex, Gale:2013da, Biswas:2019wtp}. A closer inspection of the bulk observables such as the collective flow and hadron spectra reveals the need of the inclusion of dissipative effects/transport processes in the QGP evolution~\cite{Denicol:2010xn, Baier:2006um, Baier:2007ix, Schenke:2010rr, Schenke:2011bn, Denicol:2012cn, Bhalerao:2013pza, Luzum:2008cw, Csernai:2006yk, Song:2012ua}. Besides, it has been realized that relativistic dissipative hydrodynamical modelling of the evolution of the nuclear matter is also necessary for a realistic description of various other probes such as photon production, heavy quarks, dilepton emission, and their associated flow coefficients, at LHC and RHIC~\cite{Kurian:2020orp, Paquet:2015lta, Vujanovic:2013jpa, Vujanovic:2019yih}.

Hydrodynamics is a macroscopic theory based on the fundamental laws of energy-momentum and current conservation, along with the second law of thermodynamics. On the other hand, the equation of state and the transport coefficients depend on the microscopic interactions of the medium which can be obtained from the underlying microscopic theory. Much research is devoted to explore the thermodynamic and transport properties of the QGP within the semi-classical kinetic theory~\cite{Deb:2016myz, Mitra:2017sjo, Jaiswal:2014isa, Florkowski:2015lra, Denicol:2014vaa, Kurian:2020qjr}. The earliest theoretical formulation of a relativistic theory of dissipative hydrodynamics, collectively called the Navier Stokes (NS) theory, are due to Landau-Lifshitz and Eckart, with the proper choice of fluid four-velocity~\cite{landau1987fluid, Eckart:1940zz}. However, the relativistic NS theory involves parabolic equation of motion leading to problems such as acausality and numerical instability~\cite{Hiscock:1983zz, Hiscock:1985zz, Geroch:1990bw}. The second-order theory of dissipative fluids by Muller and Israel-Stewart (IS) results in hyperbolic equation of motion and preserves causality~\cite{Muller1967329, Israel:1979wp, stewart1971non, Pu:2009fj, Brito:2020nou}. Despite the fact that the second-order theory may not guarantee the numerical stability, the IS formulation of causal theory has been quite successful in describing the QGP evolution in heavy-ion collision experiments and is an active area of current research~\cite{Denicol:2008ha, Denicol:2010xn, Denicol:2012cn, Denicol:2014vaa, Niemi:2011ix, Jaiswal:2015mxa, Mitra:2020gdk}. Recently, there have been some very interesting developments in the formulation of stable and causal theories of relativistic dissipative hydrodynamics~\cite{Kovtun:2019hdm, Hoult:2020eho, Das:2020fnr, Das:2020gtq, Biswas:2020rps}.

In order to apply the dissipative formulation of relativistic hydrodynamics to the QGP evolution, one has to ensure that thermodynamic and transport properties of QCD are incorporated via the equation of state and transport coefficients. Within the kinetic theory, the non-ideal QCD equation of state (EoS) effects can be embedded into the formulation in terms of effective quarks/antiquarks and gluons degrees of freedom. This can be done with quasiparticle description of the hot QCD matter where the medium effects are incorporated by considering thermal modification of intrinsic particle properties such as mass or fugacity. There have been few earlier attempts in the quasiparticle description of the relativistic dissipative hydrodynamics in terms of the effective mass~\cite{Goloviznin:1992ws, Peshier:1995ty, Gorenstein:1995vm, Bannur:2006js, Romatschke:2011qp, Tinti:2016bav, Alqahtani:2015qja, Alqahtani:2016rth, Alqahtani:2017jwl, Koothottil:2020riy, Czajka:2017wdo, Jaiswal:2020qmj}. In the current analysis, we have incorporated the QCD thermal medium effects within the scope of the effective fugacity quasiparticle model (EQPM)~\cite{Chandra:2011en, Chandra:2007ca, Kurian:2017yxj, Mitra:2019jld}. Further, the microscopic dynamics of the system are modelled within a consistently developed effective kinetic theory~\cite{Mitra:2018akk}. 

The estimation of transport coefficients from the underlying kinetic theory requires the knowledge of the non-equilibrium part of the quasiparticle phase-space distribution function. The two traditional approaches to determine the form of the distribution function near local thermodynamic equilibrium are Grad's $14-$moment method and the Chapman-Enskog (CE) expansion. Here we consider a simplified version of the latter approach with the relaxation time approximation (RTA) because it leads to better agreement with the distribution function obtained using microscopic Boltzmann simulations for isotropic cross-section~\cite{Plumari:2015sia}. Recently, we have analyzed the first order dissipative hydrodynamic evolution with the EQPM by employing the CE expansion method~\cite{Bhadury:2019xdf}. The focus of the present study is to extend the analysis of the system evolution to second order in space-time gradient expansion and estimate the associated transport coefficients with the effective description of hot QCD medium. We note that an earlier attempt has been made in the study of second order evolution equation of shear tensor within the EQPM using the Grad's $14-$moment method~\cite{Mitra:2019jld}. The current study consists the comprehensive analysis of second-order evolution for shear tensor and bulk viscous pressure within the CE expansion method. Further, we estimate the non-equilibrium corrections to the thermal distribution function up to second order and investigate the viscous corrections to the entropy four-current. We analyze the second-order viscous corrections to the temperature evolution and the pressure anisotropy in the $1+1-$dimensional longitudinal boost invariant expansion.

The paper is organized as follows. The theoretical formulation of the second order dissipative hydrodynamic evolution equations and the viscous corrections to the entropy four-current within the covariant kinetic theory followed by the description of longitudinal Bjorken flow are presented in section-II. Section-III contains the results and discussions of the present analysis. The summary and future outlook are presented in section-IV. 

{\bf{Notations and conventions:}} The following notations and conventions are followed in the current article. We work in natural units with $\hbar=c=k_B=1$, where $\hbar$ is the reduced Planck's constant, $c$ is the velocity of light in vacuum and $k_B$ is the Boltzmann constant. The quantity $u_\mu$ is the fluid four-velocity which takes the form $u^{\mu}=(1,0,0,0)$ in its local rest frame. The projection operator $\Delta^{\mu\nu} = g^{\mu\nu} - u^\mu u^\nu$ is orthogonal to the fluid velocity. The metric tensor has the form $g^{\mu\nu}=\text{diag}(1, -1, -1, -1)$. The quantity $\Delta^{\mu\nu}_{\alpha\beta}\equiv\frac{1}{2}(\Delta^\mu_\alpha\Delta^\nu_\beta +\Delta^\mu_\beta\Delta^\nu_\alpha)-\frac{1}{3}\Delta^{\mu\nu}\Delta_{\alpha\beta}$ defines the traceless symmetric projection operator orthogonal to the fluid velocity. The subscript $k$ in the manuscript denotes the particle species, $k=(g, q)$, where $g$ and $q$ represents gluons and quarks, respectively. The degeneracy factor for gluon and quark chosen as $g_g=N_s\times (N_c^2-1)$ and $g_q=2\times N_s\times N_c\times N_f$, where $N_s=2$ is the spin degrees of freedom, $N_f=3$ is the number of flavors, $N_c=3$ denotes the number of colors and a factor $2$ is from the antiquark contribution as we are working at the limit of zero baryon chemical potential.
%%%%%%%%%%%%%%%%%%%%%%%%%%%%%%%%%%%%%%%%%%%%%%%%%%%%%%%%%%%%%%
%
\section{Second order viscous relativistic hydrodynamics}
The description of the non-equilibrium part of the system is essential for the estimation of the dissipative hydrodynamic evolution of the QGP. We follow the recently developed covariant kinetic theory~\cite{Mitra:2018akk} within the  EQPM~\cite{Chandra:2011en, Chandra:2007ca} to describe the dynamics of the particle distribution function. The thermal QCD medium effects are incorporated in the quasiparticle description of the system. 
%
%%%%%%%%%%%%%%%%%%%%%%%%%%%%%%%%%%%%%%%%%%%%%%%%%%%%%%%%%%%%%%
\subsection{Effective covariant kinetic theory}
The relativistic transport equation quantifies the rate of change of phase-space distribution function away from the equilibrium. The first step towards the estimation of the dissipative hydrodynamic evolution of the QGP is the setting up of an effective Boltzmann equation of the system. The Boltzmann equation within the framework of the EQPM has the following form~\cite{Mitra:2018akk},
\begin{equation}\label{Beq}
\dfrac{1}{\omega_{{k}}}\Tilde{p}^{\mu}_k\partial_{\mu}f^0_k(x,\Tilde{p}_k)+F_k^{\mu} \partial^{(p)}_{\mu} f_k^0=-\dfrac{\delta f_k}{\tau_R},
\end{equation}
where $\Tilde{p}^{\mu}_k$ is the dressed quasiparticle momenta for $k$-th species. Here we have employed the RTA for the collision term with $\tau_{R}$ being the relaxation time~\cite{anderson1974relativistic}. The equilibrium distribution function appearing in the above equation is obtained within the EQPM framework and is given as
\begin{equation}\label{feq}
f^0_{q, g}=\dfrac{z_{q, g}\exp{[-\beta (u^{\mu}p_{\mu})]}}{1\pm z_{q, g}\exp{[-\beta (u^{\mu}p_{\mu})]}},
\end{equation}
where $p^{\mu}=(E,  \vec{p})$ is the bare particle momenta and $\beta=1/T$ is the inverse temperature. The quantities $z_g$, $z_q$ are the temperature-dependent effective fugacities of gluons and quarks, respectively, that encode the thermal QCD medium interactions. In the current analysis, we have considered the $(2+1)-$flavor lattice QCD equation of state~\cite{Cheng:2007jq, Borsanyi:2013bia} in order to specify the form of $z_g$, $z_q$. The quasiparticle four-momenta are related to the bare momenta through the dispersion relation,
\begin{equation}\label{dispersion reln}
\Tilde{p_k}^{\mu} = p_k^{\mu}+\delta\omega_k\, u^{\mu}, \qquad
\delta\omega_k= T^{2}\,\partial_{T} \ln(z_{k}),
\end{equation}
where $\delta\omega_k$ is the modified part of the energy dispersion of particle species $k$. The zeroth component of the particle momenta is modified as, $\Tilde{p_k}^{0}\equiv\omega_{k}=E_k+\delta\omega_k$. The force term $F_k^{\mu}=-\partial_{\nu}(\delta\omega_k u^{\nu}u^{\mu})$ is defined from the conservation of energy-momentum and particle flow, as described in the Ref.~\cite{Mitra:2018akk}.  

The energy-momentum tensor can be defined in terms of dressed momenta within the EQPM and takes the following form,
\begin{align}\label{Tmunu def KT}
T^{\mu\nu}(x)=~\sum_{k=q,g}g_k\int{d\Tilde{P}_k\,\Tilde{p}_k^{\mu}\,\Tilde{p}_k^{\nu}\,f_k(x,\Tilde{p}_{k})}+\sum_{k=q,g}\delta\omega_k\,g_k\int{d\Tilde{P}_k\,\dfrac{\langle\Tilde{p}_k^{\mu}\,\Tilde{p}_k^{\nu}\rangle}{E_k}\, f_k(x,\Tilde{p}_{k})},
\end{align}
where $f_k$ is the non-equilibrium quasiparticle phase-space distribution function. For the system close to local thermodynamic equilibrium, we have $f_k=f^0_k+\delta f_k$ with $\delta f_k/f^0_k\ll1$. Here, $\langle\Tilde{p}_k^{\mu}\Tilde{p}_k^{\nu}\rangle\equiv\frac{1}{2}(\Delta^{\mu}_{\alpha}\Delta^{\nu}_{\beta}+\Delta^{\mu}_{\beta}\Delta^{\nu}_{\alpha})\Tilde{p}_k^{\alpha}\Tilde{p}_k^{\beta}$ and $d\Tilde{P}\equiv\frac{d^3\mid\vec{\Tilde{p}}_k\mid}{(2\pi)^3\omega_{k}}$ is the phase space factor. In general, the energy-momentum tensor can also be decomposed in terms of hydrodynamic degrees of freedom as,
\begin{equation}\label{Tmunu def hydro}
T^{\mu\nu}=\varepsilon u^{\mu} u^{\nu} - (P+\Pi) \Delta^{\mu\nu} + \pi^{\mu\nu},
\end{equation} 
in which $\varepsilon$ and $P$ are the energy density and pressure of the system, respectively. The dissipative quantities in the Eq.~(\ref{Tmunu def hydro}) are the shear stress tensor $\pi^{\mu\nu}$ and bulk viscous pressure $\Pi$, respectively. Note that the above expression for energy-momentum tensor is written for fluid four-velocity defined in the Landau frame~\cite{landau1987fluid}. 

The projection of energy-momentum conservation $\partial_{\mu}T^{\mu\nu}=0$, along and orthogonal to $u^{\mu}$, along with the thermodynamic identities give the evolution equation of $\varepsilon$ and $u^{\mu}$ and have the following form,
\begin{align}
\dot{\varepsilon}+(\varepsilon+P+\Pi)\theta-\pi^{\mu\nu}\sigma_{\mu\nu}&=0,\label{hydroeq1}\\
(\varepsilon+P+\Pi)\dot{u}_{\alpha}-\nabla^{\alpha}(P+\Pi)+\Delta^\alpha_\nu\partial_{\mu}\pi^{\mu\nu}&=0,\label{hydroeq2} 
\end{align}
where, $\theta\equiv\partial_{\mu}u^{\mu}$ is the scalar expansion and $\sigma^{\mu\nu}\equiv\Delta^{\mu\nu}_{\alpha\beta}\nabla^{\alpha}u^{\beta}$ is the shear stress tensor. From Eqs.~(\ref{hydroeq1}) and~(\ref{hydroeq2}), we can obtain the derivatives of $\beta$ which have the following form,
\begin{align}\label{beta derivative}
&\dot{\beta} = \beta c_s^2\left(\theta + \frac{\Pi\theta-\pi^{\mu\nu}\sigma_{\mu\nu}}{(\varepsilon+P)}\right),
&&\nabla^{\alpha}\beta = -\beta \Big(\dot{u}^{\alpha}  + \frac{\Pi\dot{u}^{\alpha} - \nabla^{\alpha}\Pi + \Delta^{\alpha}_{\nu} \partial_{\mu}\pi^{\mu\nu}}{(\varepsilon+P)}\Big),
\end{align} 
where $c_s^2=\frac{dP}{d\epsilon}$ is the squared speed of sound in the QGP. While deriving Eq.~\eqref{beta derivative}, we observe that with zero chemical potential, the above form of $\dot{\beta}$ and $\nabla^\alpha \beta$ are universal as long as the thermodynamic relation, $dP/dT =\beta (\epsilon + P)$ is satisfied. In other words, the changes due to the mean fields are absorbed into $c_s^2$ and thus do not affect the form of $\dot{\beta}$ or $\nabla^\alpha\beta$. The shear stress tensor $\pi^{\mu\nu}$ and  bulk viscous pressure $\Pi$ can be expressed in terms of $\delta f$ within the EQPM as~\cite{Mitra:2018akk},
\begin{align}
\pi^{\mu\nu}&=\sum_{k=q,g} g_k\Delta^{\mu\nu}_{\alpha\beta}\int{d\Tilde{P}_k~   \Tilde{p}_k^{\alpha}\Tilde{p}_k^{\beta}\delta f_k}+\sum_{k=q,g} \delta \omega_k g_k\Delta^{\mu\nu}_{\alpha\beta}\int{d\Tilde{P}_k~   \Tilde{p}_k^{\alpha}\Tilde{p}_k^{\beta}\dfrac{1}{E_k}\delta f_k}, \label{pimunu def KT}\\
\Pi&=-\dfrac{1}{3}\sum_{k=q,g} g_k\Delta_{\alpha\beta}\int{d\Tilde{P}_k~    \Tilde{p}_k^{\alpha}\Tilde{p}_k^{\beta}\delta f_k}-\dfrac{1}{3}\sum_{k=q,g} \delta \omega_k g_k\Delta_{\alpha\beta}\int{d\Tilde{P}_k~   \Tilde{p}_k^{\alpha}\Tilde{p}_k^{\beta}\dfrac{1}{E_k}\delta f_k}. \label{Pi def KT}
\end{align}
To make further progress, we need to obtain $\delta f_k$ in order to describe the transport coefficients in the viscous evolution of the system.

%%%%%%%%%%%%%%%%%%%%%%%%%%%%%%%%%%%%%%%%%%%%%%%%%%%%5
\subsection{Second order viscous evolution equations}
In the present analysis, we obtain non-equilibrium corrections to the distribution function by employing an iterative CE like solution of the Boltzmann equation, Eq.~(\ref{Beq}), in RTA~\cite{Jaiswal:2013npa, Jaiswal:2013vta}. The first-order gradient correction to distribution functions for quarks and gluons within the effective kinetic theory takes the following form,
\begin{align}\label{del f}
\delta f_k^{(1)} &= \tau_R\bigg[ \Tilde{p}_k^\gamma\partial_\gamma \beta + \frac{\beta\, \Tilde{p}_k^\gamma\, \Tilde{p}_k^\phi}{u\!\cdot\!\Tilde{p}_k}\partial_\gamma u_\phi - \beta \theta (\delta\omega_k) - \beta \dot{\beta} \left(\frac{\partial(\delta\omega_k)}{\partial\beta}\right)\bigg] f_{k}^0\bar{f}_{k}^0,
\end{align}
where $\bar{f}_{k}^0=1-af_{k}^0$ with $a=-1$ and $+1$ for gluons and quarks, respectively. Employing the first order equation and considering the relaxation time $\tau_{R}$ to be independent of four-momenta, Eqs.~(\ref{pimunu def KT}) and Eqs.~(\ref{Pi def KT}) reduce to,
\begin{align}\label{NS eqns}
&\pi^{\mu\nu}=2\,\tau_R\,\beta_\pi\,\sigma^{\mu\nu},  &&\Pi=-\tau_R\,\beta_\Pi\,\theta.
\end{align}
The above equation is the relativistic generalization of the Navier-Stokes equation. The coefficients $\beta_\pi$ and $\beta_\Pi$ have been estimated in terms of the thermodynamic integrals $\Tilde{J}^{(r)}_{k~ nm}$ and $\Tilde{L}^{(r)}_{k~ nm}$ as, 
\begin{align}
\beta_\pi =&\, \beta\sum_{k=q,g}  \bigg[\Tilde{J}^{(1)}_{k~42}+\delta\omega_k\Tilde{L}^{(1)}_{k~42}\bigg],\label{beta pi}\\
\beta_\Pi =&\, \beta \!\! \sum_{k=q,g} \! \bigg[\! c_s^2 \! \bigg(\! \Tilde{J}_{k~31}^{(0)} \!+\! \delta\omega_k \Tilde{L}_{k~31}^{(0)} \!-\! \left(\!\frac{\partial(\delta\omega_k)}{\partial\beta}\!\right) \!\! \left(\Tilde{J}_{k~21}^{(0)} + \delta\omega_k \Tilde{L}_{k~21}^{(0)}\right)\!\!\!\bigg) \!+\! \frac{5}{3}\bigg(\! \Tilde{J}_{k~42}^{(1)} + \delta\omega_k \Tilde{L}_{k~42}^{(1)} \!\bigg)\nonumber\\
&- \delta\omega_k \left(\Tilde{J}_{k~21}^{(0)} + \delta\omega_k \Tilde{L}_{k~21}^{(0)}\right) \!\bigg]\label{beta Pi}.
\end{align}
The thermodynamic integrals $\Tilde{J}^{(r)}_{k~nm}$ and $\Tilde{L}^{(r)}_{k~nm}$ are defined as,
\begin{align}
\Tilde{J}^{(r)}_{k~nm}&=\dfrac{g_{k}}{2\pi^2}\frac{(-1)^m}{(2m+1)!!}\int_{0}^\infty{d\mid\vec{\Tilde{p}}_k\mid}~{\big(u.\Tilde{p}_k\big)^{n-2m-r-1}}\big(\mid\vec{\Tilde{p}}_k\mid\big)^{2m+2}f^0_k\bar{f}^0_k, \label{Jnmr}\\
\Tilde{L}^{(r)}_{k~nm}&=\dfrac{g_{k}}{2\pi^2}\frac{(-1)^m}{(2m+1)!!}\int_{0}^\infty{d\mid\vec{\Tilde{p}}_k\mid}~\dfrac{\big(u.\Tilde{p}_k\big)^{n-2m-r-1}}{E_k}\big(\mid\vec{\Tilde{p}}_k\mid\big)^{2m+2}f^0_k\bar{f}^0_k.\label{Lmnr}
\end{align}
These integral coefficients can be expressed in terms of polylogarithm functions and are given in the Appendix~\ref{A}. The first-order transport coefficients, shear viscosity $\eta = \tau_R \beta_\pi$ and bulk viscosity $\zeta = \tau_R \beta_\Pi$, within the effective kinetic theory have been calculated by comparing Eq.~(\ref{NS eqns}) with the relativistic Navier-Stokes equations as discussed in Ref.~\cite{Bhadury:2019xdf}. 

To obtain the second order hydrodynamic evolution equations for the shear stress tensor and the bulk viscous pressure, we adopt the methodology followed in Refs.~\cite{Denicol:2010xn, Jaiswal:2014isa}. The co-moving derivative of $\pi^{\mu\nu}$ and $\Pi$ within the covariant kinetic theory take the following forms,
\begin{align}
\dot{\pi}^{\langle\mu\nu\rangle} &= \sum_{k=q,g}g_k\, \Delta^{\mu\nu}_{\alpha\beta}\, D\left(\int\mathrm{d\Tilde{P}_k}~\Tilde{p}_k^\alpha \Tilde{p}_k^\beta \delta f_{k} + (\delta\omega_k) \int\frac{\mathrm{d\Tilde{P}_k}}{ E_{k}} \Tilde{p}_k^\alpha \Tilde{p}_k^\beta \delta f_{k}\right),\label{shear evolution def KT}\\
\dot{\Pi} &= -\frac{1}{3}\sum_{k=q,g} g_k \Delta_{\alpha\beta}\, D \Bigg(\int\mathrm{d\Tilde{P}_k}\Tilde{p}_k^\alpha \Tilde{p}_k^\beta \delta f_{k} + (\delta\omega_k) \int\frac{\mathrm{d\Tilde{P}_k}}{ E_{k}} \Tilde{p}^\alpha \Tilde{p}^\beta \delta f_{k}\Bigg),\label{bulk evolution def KT}
\end{align}
where we have employed the notation ${X}^{\langle\mu\nu\rangle} \equiv \Delta^{\mu\nu}_{\alpha\beta} X^{\alpha\beta}$ and $D(X) \equiv \dot{X}$. The co-moving derivative of the non-equilibrium part of distribution function i.e. $\delta \dot{f}_{k}$ can be obtained from Eq.~(\ref{Beq}) and takes the following form,
\begin{align}\label{del f evolution}
\delta\dot{f}_{k} &= - \dot{f}^0_{k} - \frac{\delta f_{k}}{\tau_R} - \frac{1}{(u\cdot \Tilde{p}_k)} \Tilde{p}_k^\gamma \nabla_\gamma f_{k} + \Big[\delta \dot{\omega}_k u^\gamma + \delta \omega_k \theta u^\gamma + \delta \omega_k \dot{u}^\gamma\Big] \partial_\gamma^{(\Tilde{p}_k)} f_{k}.  
\end{align}
Substituting $\delta\dot{f}_{k}$ in the Eqs.~(\ref{shear evolution def KT}) and~(\ref{bulk evolution def KT}) along with employing the Eqs.~(\ref{beta derivative}) and~(\ref{del f}), we finally obtain the second order evolution equations for shear tensor and bulk viscous pressure as,
\begin{align}\label{shear evolution def hydro}
\dot{\pi}^{\langle\mu\nu\rangle} + \frac{\pi^{\mu\nu}}{\tau_R} =&~ 2 \beta_{\pi} \sigma^{\mu\nu} + 2 \pi_{\phi}^{\langle\mu} \omega^{\nu\rangle\phi} - \delta_{\pi\pi} \pi^{\mu\nu} \theta - \tau_{\pi\pi} \pi_{\phi}^{\langle\mu} \sigma^{\nu\rangle \phi} + \lambda_{\pi\Pi} \Pi \sigma^{\mu\nu},\\
\dot{\Pi} + \frac{\Pi}{\tau_R} =&~ - \beta_{\Pi} \theta - \delta_{\Pi\Pi} \Pi \theta + \lambda_{\Pi\pi} \pi^{\mu\nu} \sigma_{\mu\nu},\label{bulk evolution def hydro}
\end{align}
with $\omega^{\mu\nu} = \frac{1}{2} (\nabla^{\mu} u^{\nu} - \nabla^{\nu} u^{\mu})$ as the vorticity tensor. Note that, as a consequence of RTA, one obtains a single time scale to describe the relaxation of shear and bulk viscous evolution, $i.e$., $\tau_R=\tau_{\pi}=\tau_{\Pi}$. 

The second order transport coefficients appearing in the viscous evolution Eqs.~\eqref{shear evolution def hydro} and \eqref{bulk evolution def hydro} are obtained as
\begin{align}
\delta_{\pi\pi}&=\dfrac{5}{3}+\dfrac{\beta}{\beta_{\pi}}\sum_{k=q,g}\bigg[\dfrac{7}{3}\Tilde{J}^{(3)}_{k~63}+{\delta{\omega}_k}\Big(\dfrac{7}{3}\Tilde{L}^{(3)}_{k~63}-\dfrac{7}{6}\xi_k+\dfrac{1}{2}\Gamma_k\Big)\bigg],\label{delta pipi}\\
\tau_{\pi\pi}&= 2 + \frac{\beta}{\beta_{\pi}} \sum_{k=q,g} \bigg[4\Big(\Tilde{J}^{(3)}_{k~63}+\delta\omega_k\Tilde{L}^{(3)}_{k~63}\Big)-{2\delta\omega_k }\xi_k\bigg],\label{tau pipi}\\
\lambda_{\pi\Pi}&= \dfrac{\beta c_s^2}{\beta_{\Pi}}\sum_{k=q,g}\bigg[\Tilde{J}^{(1)}_{k~42}+\Tilde{J}^{(0)}_{k~31}+\delta\omega_k\Big(\Tilde{L}^{(1)}_{k~42}-\Tilde{J}^{(0)}_{k~21}+\Tilde{L}^{(0)}_{k~31}-\delta\omega_k\Tilde{L}^{(0)}_{k~21}\Big) +\beta\dfrac{\partial\delta\omega_k}{\partial\beta}\Big(2\xi_k+\Gamma_k\nonumber\\
&~~-2\delta\omega_k\Tilde{L}^{(1)}_{k~42}\Big)\!\bigg]+\dfrac{\beta}{\beta_{\Pi}}\sum_{k=q,g}\!\!\bigg[\dfrac{14}{3}\Tilde{J}^{(3)}_{k~63}+\dfrac{10}{3}\Tilde{J}^{(1)}_{k~42}+\delta\omega_k\Big(\dfrac{14}{3}\Tilde{L}^{(3)}_{k~63}+\dfrac{10}{3}\Tilde{L}^{(1)}_{k~42}-\dfrac{7}{3}\xi_k+\Gamma_k\Big)\!\bigg],\label{lambda piPi}\\
\lambda_{\Pi\pi} &= \dfrac{\beta}{3\beta_{\pi}}\sum_{k=q,g}\bigg[7\Tilde{J}^{(3)}_{k~63}+2\Tilde{J}^{(2)}_{k~52}+\delta\omega_k\Big(7\Tilde{L}^{(3)}_{k~63}+2\Tilde{L}^{(2)}_{k~52}-2\xi_k\Big)\bigg]-c_s^2,\label{lambda Pipi}\\
\delta_{\Pi\Pi}&=\frac{\beta}{\beta_{\Pi}} \!\! \sum_{k=q,g} \!\! \bigg[ \!\!-\! \frac{5}{9}\lambda_{0k} \!-\! \delta\omega_k\lambda_{1k} \!+\! \!\left(\!\frac{\partial\delta\omega_k}{\partial\beta}\!\right)\!\lambda_{2k} \!-\! (\delta\omega_k)^2 \lambda_{3k}\!+\! \delta\omega_k \!\left(\!\frac{\partial\delta\omega_k}{\partial\beta}\!\right)\!\! \lambda_{4k} \!-\! \left(\!\frac{\partial\delta\omega_k}{\partial\beta}\!\right)^2 \!\!\!\lambda_{5k}\!\bigg] \!-\! c_s^2,\label{delta PiPi}
\end{align}
where 
\begin{align}
\xi_k =&~ \Tilde{J}^{(2)}_{k~42}+\delta\omega_k\Tilde{L}^{(2)}_{k~42},\\
\Gamma_k =&~ \Tilde{J}^{(0)}_{k~21}-\beta \Tilde{M}^{(1)}_{k~42} +\delta\omega_k\Big(\Tilde{L}^{(0)}_{k~21}-\Tilde{J}^{(1)}_{k~21}-\beta \Tilde{N}^{(1)}_{k~42} \Big),\\
\lambda_{0k} =&~ =\Big(\Tilde{J}^{(1)}_{k~42}+\Tilde{J}^{(0)}_{k~31}+\delta\omega_k(\Tilde{L}^{(1)}_{k~42}+\Tilde{L}^{(0)}_{k~31})\Big)(1-3c_s^2),\\
\lambda_{1k} =&~ \Big(\dfrac{8}{3}\Tilde{J}^{(0)}_{k~21}-\beta \Tilde{M}^{(0)}_{k~31}\Big)c_s^2+\dfrac{25}{9}\Tilde{J}^{(2)}_{k~42}-\dfrac{5}{3}\Tilde{J}^{(1)}_{k~31}-\dfrac{5}{3}\beta\Tilde{M}^{(1)}_{k~42},\\
\lambda_{2k} =&~ \dfrac{5}{3}\Big(\Tilde{J}^{(1)}_{k~31}+\Tilde{J}^{(2)}_{k~42}+\beta\Tilde{M}^{(1)}_{k~42}-\Tilde{L}^{(1)}_{k~42}\Big)\beta c_s^2+\Tilde{M}^{(0)}_{k~31}\beta^2(c_s^2)^2,\\
\lambda_{3k} =&~ \dfrac{5}{3}\Tilde{J}^{(1)}_{k~21}-\beta\Tilde{M}^{(0)}_{k~21}+\Big(\dfrac{8}{3}\Tilde{L}^{(0)}_{k~21}-\beta\Tilde{N}^{(0)}_{k~31}\Big)c_s^2+\dfrac{25}{9}\Tilde{L}^{(2)}_{k~42}-\dfrac{5}{3}\Tilde{L}^{(1)}_{k~31}-\dfrac{5}{3}\beta\Tilde{N}^{(1)}_{k~42},\\
\lambda_{4k} =&~ \Big(\dfrac{1}{3}\Tilde{J}^{(1)}_{k~21}+2\beta\Tilde{M}^{(0)}_{k~21}\Big)\beta c_s^2+\Tilde{N}^{(0)}_{k~31}\beta^2 (c_s^2)^2+\dfrac{5}{3}\Big(\dfrac{3}{5}\Tilde{L}^{(0)}_{k~21}+\Tilde{L}^{(1)}_{k~31}+\Tilde{L}^{(2)}_{k~42}+\beta\Tilde{N}^{(1)}_{k~42}\Big)\beta c_s^2,\\
\lambda_{5k} =&~ \Big(\Tilde{J}^{(1)}_{k~21}-\beta\Tilde{M}^{(0)}_{k~21}+\beta^{-1}\Tilde{L}^{(0)}_{k~31}\Big)\beta^2 (c_s^2)^2.
\end{align}
In the above expressions, the integral coefficients $\Tilde{M}^{(r)}_{k~nm}$ and $\Tilde{N}^{(r)}_{k~nm}$ are defined as,
\begin{align}
\Tilde{M}^{(r)}_{k~nm} =&~ \dfrac{g_{k}}{2\pi^2}\frac{(-1)^m}{(2m+1)!!}\int_{0}^\infty{d\mid\vec{\Tilde{p}}_k\mid}~{\big(u.\Tilde{p}_k\big)^{n-2m-r-1}}\big(\mid\vec{\Tilde{p}}_k\mid\big)^{2m+2}(\bar{f}^0_k-af^0_k)f^0_k\bar{f}^0_k, \label{Mmnr}\\
\Tilde{N}^{(r)}_{k~nm} =&~ \dfrac{g_{k}}{2\pi^2}\frac{(-1)^m}{(2m+1)!!}\int_{0}^\infty{d\mid\vec{\Tilde{p}}_k\mid}~\dfrac{{\big(u.\Tilde{p}_k\big)^{n-2m-r-1}}}{E_k}\big(\mid\vec{\Tilde{p}}_k\mid\big)^{2m+2}(\bar{f}^0_k-af^0_k)f^0_k\bar{f}^0_k. \label{Nmnr}
\end{align}
The thermodynamic integrals $\Tilde{J}^{(r)}_{k~nm},~\Tilde{L}^{(r)}_{k~nm},~\Tilde{M}^{(r)}_{k~nm}$, and $\Tilde{N}^{(r)}_{k~nm}$, appearing in the above expressions, can be expressed in terms of polylogarithm functions and are given in the Appendix~\ref{A}. It is important to note that the transport coefficients described in the Eqs.~(\ref{delta pipi})-(\ref{delta PiPi}) reduce exactly to the results in the Ref.~\cite{Florkowski:2015lra} in the non-interacting ideal case where $z_k\rightarrow 1$ and $\delta\omega_k\rightarrow 0$. 
%%%%%%%%%%%%%%%%%%%%%%%%%%%%%%%%%%%%%%%%%%%%%%%%%%%%%%%%%%%%%%%
%
\subsection{Second-order viscous corrections to entropy-four current}
Another approach to obtaining equations of relativistic dissipative hydrodynamic follow from the second law of thermodynamics~\cite{Bhattacharyya:2012nq, Bhattacharya:2019qal}. This requires the condition of positive definite local entropy generation in the medium. In a relativistic system, entropy generation can be realized from the divergence of entropy four-current. In the following, we construct the entropy four-current in terms of hydrodynamic quantities within the EQPM framework. The entropy four-current can be defined from the Boltzmann H-function as,
\begin{equation}\label{entr_curr}
S^{\mu} = -\sum_{k}g_k\int\mathrm{d\Tilde{P}}_k\,\Tilde{p}^{\mu}_k\left(f_k\ln{f_k}+a\bar{f}_k\ln{\bar{f}_k}\right).
\end{equation}
For a system near to the local thermodynamic equilibrium, the distribution function can be written as $f_k=f_k^0+f_k^0 \bar{f}^0_k\phi_k$ where $\phi_k\ll1$. Substituting in the above equation and expanding in powers of $\phi_k$ up to second order, we obtain
\begin{align}\label{entropy current}
S^\mu =&~ s_0 u^\mu - \sum_k g_k\! \int\mathrm{d\Tilde{P}_k}\, \Tilde{p}_k^\mu \left[f_k^0\, \bar{f}_k^0\, \ln{\left(f_k^0/ \bar{f}_k^0\right)}\, \phi_k + \frac{1}{2}\,f_k^0 \bar{f}_k^0\, \phi_k^2\right]+\mathcal{O}(\phi_k^3),
\end{align}
where $s_0=\beta(\varepsilon+P)$ is the equilibrium entropy density. Employing the EQPM distribution function, Eq.~\eqref{feq}, in Eq.~\eqref{entropy current}, we obtain
\begin{align}\label{entropy current def KT EQPM}
S^\mu =&~ s_0 u^\mu +\sum_k g_k \ln{\left(z_{1k}\right)} \! \int\mathrm{d\Tilde{P}_k}\, \Tilde{p}_k^\mu\, f_k^0 \bar{f}_k^0 \phi_k - \sum_k g_k\! \int\mathrm{d\Tilde{P}_k}\, \Tilde{p}_k^\mu \frac{1}{2}\, f_k^0 \bar{f}_k^0\, \phi_k^2,
\end{align}
where $z_{1k} = z_k \exp{[\beta(\delta\omega_k)]}$. 

In Eq.~\eqref{entropy current}, we see that the first-order viscous correction to the entropy current is purely due to quasiparticle excitation because the term is proportional to $\ln z_{1k}$. Note that, within the usual kinetic theory with zero chemical potential, first order correction to entropy current does not arise~\cite{Jaiswal:2013fc}. This limit is easily recovered in Eq.~(\ref{entropy current def KT EQPM}) by setting $z_k=1$ which corresponds to the limit of ideal EoS. We find that the first-order term in the Eq.~(\ref{entropy current def KT EQPM}) vanishes, and the second-order derivative expansion terms contribute to leading order non-equilibrium corrections to the entropy four-current~\cite{Chattopadhyay:2014lya}. In Ref.~\cite{Bhadury:2019xdf}, we have realized the non-zero first order bulk viscous correction to the entropy current within the EQPM description. To obtain the second order viscous corrections to the entropy four-current, the knowledge of the non-equilibrium parts of the distribution function up to second order in space-time derivatives of hydrodynamic variables are required. To this end, we first write the non-equilibrium part of the distribution function as a sum of first and second order corrections, $\phi_k=\phi_{1k}+\phi_{2k}$. Invoking the evolution equations Eq.~(\ref{shear evolution def hydro}) and Eq.~(\ref{bulk evolution def hydro}), the first and second order viscous corrections to the distribution function are calculated to be,
\begin{align}
\phi_{1k} =&~ {C}_{1k}\, \Tilde{p}_k^\alpha\, \Tilde{p}_k^\beta\, \pi_{\alpha\beta} + {C}_{2k} \Pi, \label{phi 1}\\
\phi_{2k} =&~ {C}_{3k}\, \Tilde{p}_k^\alpha (\dot{u}^\beta \pi_{\alpha\beta}) \!+\! {C}_{4k} \Tilde{p}_k^\alpha\, \Tilde{p}_k^\beta\, \Tilde{p}^\mu_k (\dot{u}_\mu \pi_{\alpha\beta}) \!+\! {C}_{5k}\, \Tilde{p}^\alpha_k (\nabla^\beta \pi_{\alpha\beta}) \!+\! {C}_{6k}\, \Tilde{p}_k^\alpha \Tilde{p}_k^\beta \Tilde{p}^\mu_k (\nabla_\mu \pi_{\alpha\beta}) \!+\! {C}_{7k} (\pi^{\alpha\beta} \pi_{\alpha\beta}) \nonumber\\
&+ {C}_{8k}\, \Tilde{p}_k^\alpha\, \Tilde{p}^\beta_k (\pi_\beta^\gamma\, \pi_{\alpha\gamma}) + {C}_{9k}\, \Tilde{p}_k^\alpha\, \Tilde{p}_k^\beta\, \Tilde{p}^\mu_k \Tilde{p}_k^\gamma (\pi_{\mu\gamma} \pi_{\alpha\beta}) + {C}_{10k}\, \Tilde{p}_k^\alpha\, \Tilde{p}^\beta_k (\pi_{\alpha}^{\gamma}\, \omega_{\beta\gamma}) + {C}_{11k} \Tilde{p}^\alpha_k\, \Tilde{p}_k^\beta (\pi_{\alpha\beta} \Pi) \nonumber\\
&+ {C}_{12k}\, \Tilde{p}^\alpha_k (\dot{u}_\alpha\, \Pi) + {C}_{13k}\, \Tilde{p}^\alpha_k (\nabla_\alpha \Pi)+ {C}_{14k}\, (\Pi^2), \label{phi 2}
\end{align}
where the coefficients $C_{i}$ $(i=1, 2, ....,  15)$ for the $k-$th particle species are given in the Appendix~\ref{B}.

The non-equilibrium corrections to the distribution function as obtained in Eq.~(\ref{phi 1}) and Eq.~(\ref{phi 2}) reduce back to the form in Ref.~\cite{Chattopadhyay:2014lya} in the ultra-relativistic limit $z_k\rightarrow 1$. Substituting $\phi_{1k}$ and $\phi_{2k}$ in Eq.~(\ref{entropy current def KT EQPM}), the entropy current can be expressed as,
\begin{align}\label{entropy current EQPM}
S^\mu = s u^\mu + \Phi^\mu,
\end{align}
where $s \equiv u_\mu S^\mu$ is the entropy density evaluated by projecting the entropy four-current along the fluid velocity and $\Phi^\mu \equiv \Delta^\mu_\alpha S^\alpha$ is the entropy flux which is the space-like component obtained by projecting the entropy four-current orthogonal to fluid velocity. We obtain the entropy density and the entropy flux as follows,
\begin{align}
s =&~ s_0 + \Lambda_\Pi \Pi + \Lambda_{\pi\pi}\, (\pi_{\alpha\beta} \pi^{\alpha\beta}) + \Lambda_{\Pi\Pi}\, \Pi^2, \label{entropy density}\\
\Phi^\mu =&~ \Lambda_1 (\Delta^\mu_\alpha \dot{u}_\beta \pi^{\alpha\beta}) + \Lambda_2 \left( \Delta^\mu_\alpha \nabla_\beta \pi^{\alpha\beta}\right) + \Lambda_3 (\dot{u}^\mu\, \Pi) + \Lambda_4 (\nabla^\mu \Pi).\label{entropy flux}
\end{align}
The $\Lambda$ coefficients in Eqs.~\eqref{entropy density} and \eqref{entropy flux} are provided in Appendix~\ref{B}. It is important to note that in the usual kinetic theory with vanishing chemical potential, the entropy flux, $\Phi^\mu$, does not appear up to second-order gradient corrections. On the other hand, we find non-vanishing $\Phi^\mu$ in the present calculation which appears purely due to consideration of effective fugacity. This is apparent from the expressions of $\Lambda_i$ ($i=1,2,3,4$) in Appendix~\ref{B} where these coefficients are proportional to $\ln z_{1k}$ and vanishes in the limit $z_k\rightarrow1$.

%%%%%%%%%%%%%%%%%%%%%%%%%%%%%%%%%%%%%%%%%%%%%%%%%%%%%%%%%%%%%
\begin{figure*}
\centering
\includegraphics[scale=0.435]{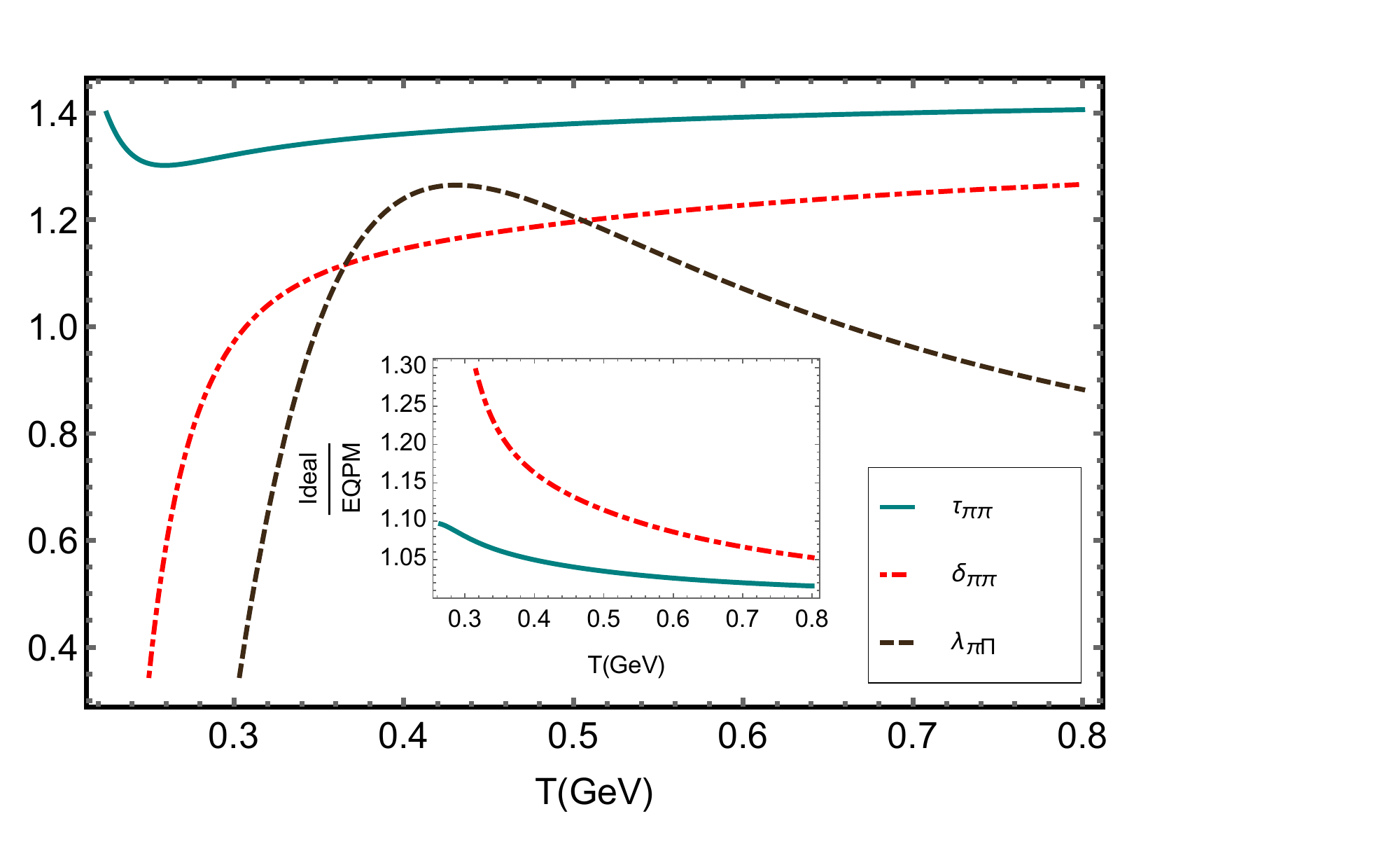}
\hspace{-.5 cm}
\includegraphics[scale=0.375]{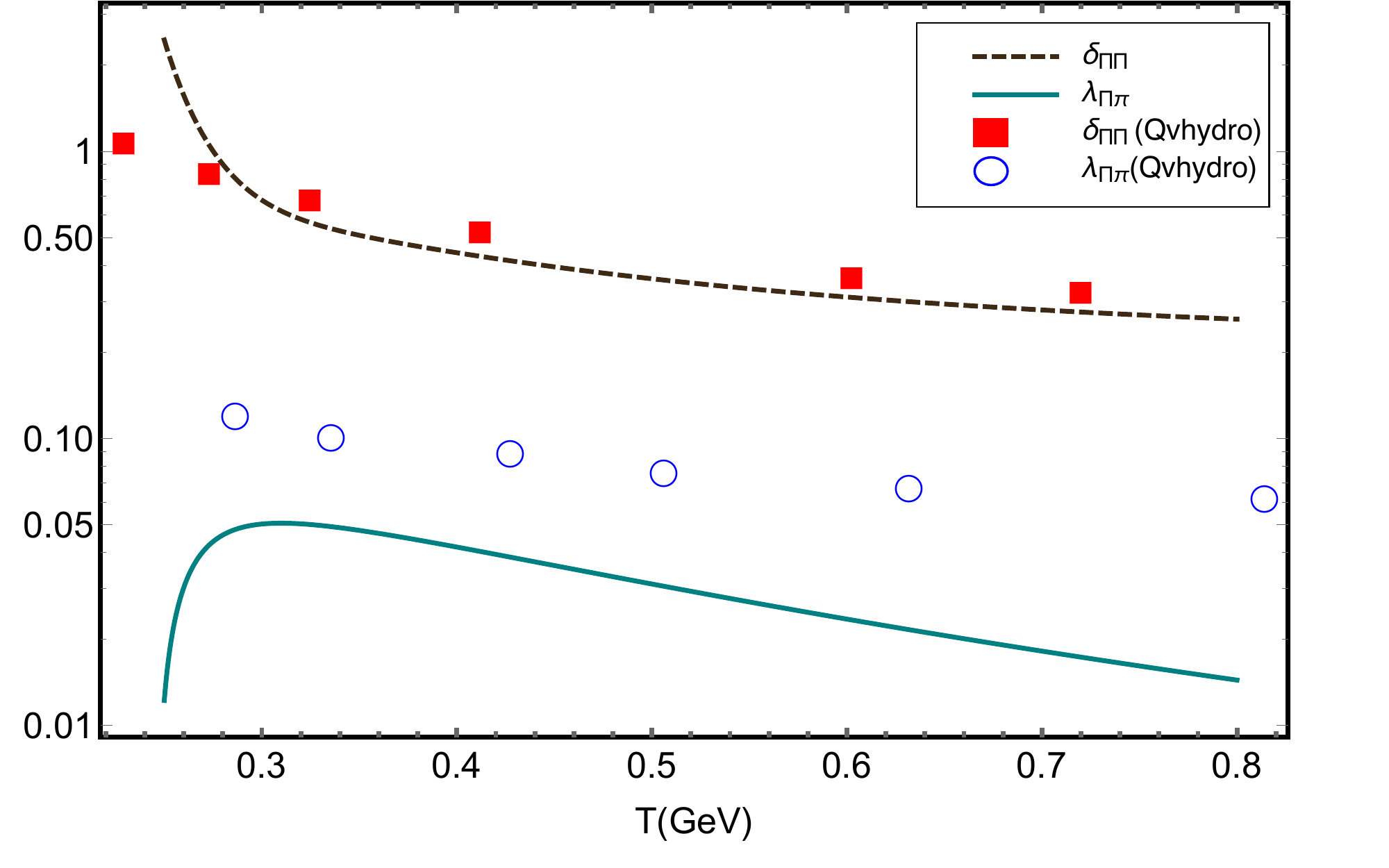}
\caption{Temperature behaviour of shear second-order transport coefficients within the present EQPM formulation and the comparison of the EQPM results with that obtained with ideal EoS (left panel). Temperature dependence of bulk second-order transport coefficients (right panel). The results are compared with effective mass quasiparticle hydrodynamic formulation~\cite{Tinti:2016bav}.}
\label{f1}
\end{figure*} 
%%%%%%%%%%%%%%%%%%%%%%%%%%%%%%%%%%%%%%%%%%%%%%%%%%%%%%%%%%%%

\subsection{Boost invariant Bjorken expansion}
The EoS effects from the EQPM formulation of the second-order viscous hydrodynamic equations to study the evolution of QGP can be quantified by employing the Bjorken's prescription for transversely homogeneous and purely longitudinal boost-invariant expansion~\cite{Bjorken:1982qr}. In terms of Milne coordinates $(\tau,x,y,\eta_s)$, with $\tau=\sqrt{t^2-z^2}$ and $\eta_s=\tanh^{-1}(z/t)$, fluid velocity takes the form, $u^{\mu}=(1,0,0,0)$, in which the metric tensor is given by $g^{\mu\nu}=(1,-1,-1,-1/\tau^2)$. Employing the Milne coordinate system, the evolution equation of energy density \textit{i.e.} Eq.~(\ref{hydroeq1}), together with the viscous evolution, Eqs.~(\ref{shear evolution def hydro}) and~(\ref{bulk evolution def hydro}), reduce to
\begin{align}
\frac{d\varepsilon}{d\tau} =&~ -\frac{1}{\tau}\bigg({\varepsilon+P+\Pi-\pi}\bigg),\label{energy Bjorken evolution}\\
\frac{d\pi}{d\tau}+\frac{\pi}{\tau_{\pi}} =&~ \dfrac{4}{3} \frac{\beta_{\pi}}{\tau}-\bigg(\frac{1}{3}\tau_{\pi\pi}+\delta_{\pi\pi}\bigg)\frac{\pi}{\tau}+\frac{2}{3}\lambda_{\pi\Pi} \frac{\Pi}{\tau}, \label{shear Bjorken evolution}\\
\dfrac{d\Pi}{d\tau}+\frac{\Pi}{\tau_{\Pi}} =&~ -\frac{\beta_{\Pi}}{\tau}-\delta_{\Pi\Pi}\frac{\Pi}{\tau}+\lambda_{\Pi\pi}\frac{\pi}{\tau},\label{bulk Bjorken evolution}
\end{align}
where $\pi\equiv-\tau^2\pi^{\eta_s\eta_s}$. The transport coefficients appearing in the shear and bulk evolution equations are described in the Eq.~(\ref{delta pipi})-(\ref{delta PiPi}). Note that the term with the vorticity tensor $2\pi_{\phi}^{\langle\mu}\omega^{\nu \rangle\phi}$ vanishes in the evolution equations and has no effect on the dynamics of the fluid. We numerically solve the simultaneous Eqs.~(\ref{energy Bjorken evolution})-(\ref{bulk Bjorken evolution}) to study the evolution of viscous QGP with initial temperature $T_0 = 600$ MeV at the initial proper time $\tau_0 = 0.25$~fm/c corresponding to the LHC initial conditions. The parameters $z_k$ and $\delta\omega_k$ are obtained by imposing the lattice QCD equation of state. We consider relaxation time to be constant and equal for both bulk and shear parts, $\tau_\pi =\tau_\Pi=\tau_R=0.25$~fm/c. The pressure anisotropy in the medium can be written as $P_L/P_T\equiv (P+\Pi-\pi)/(P+\Pi+\pi/2)$, where $P_L$ and $P_T$ are the longitudinal pressure and transverse pressure, respectively. Further, with these conditions we can study the system response due to the viscous flow by investigating the Reynolds number associated with shear and bulk viscous pressure, $R^{-1}_\pi= \frac{\sqrt{\pi_{\mu\nu}\pi^{\mu\nu}}}{P}$ and $R^{-1}_\Pi=  \frac{\lvert\,\Pi\,\rvert}{P}$, respectively~\cite{Jeon:2015dfa, Denicol:2012cn, Muronga:2003ta}. 
%
%%%%%%%%%%%%%%%%%%%%%%%%%%%%%%%%%%%%%%%%%%%%%%%%%%%%%%%%%%%%%%%
\section{Results and discussions}

%%%%%%%%%%%%%%%%%%%%%%%%%%%%%%%%%%%%%%%%%%%%%%%%%%%%%%%%%%
\begin{figure}
\centering
\includegraphics[scale=0.46]{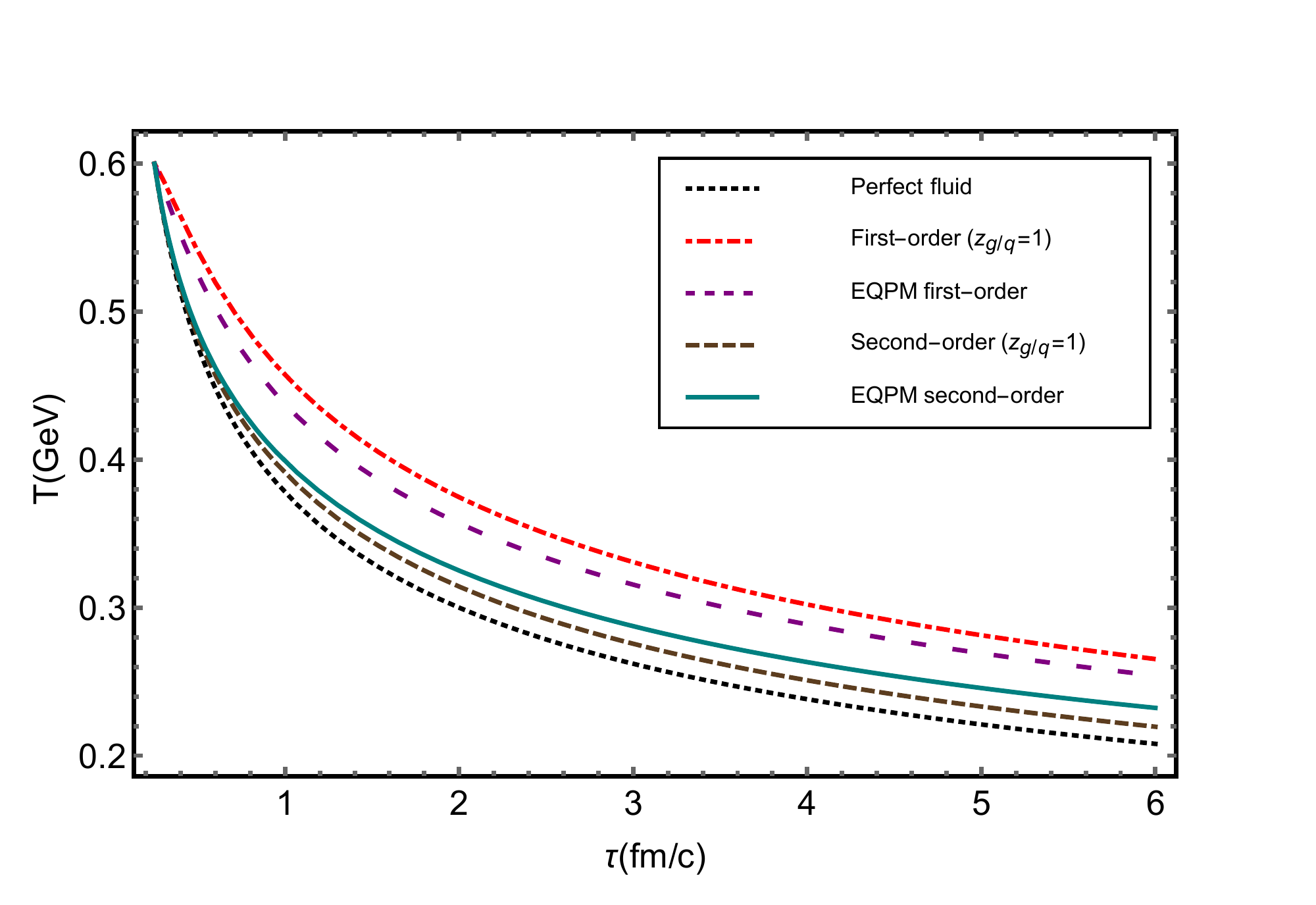}
\caption{ Proper time evolution of temperature in the medium with initial temperature $T_0 = 600$ MeV at the initial proper time $\tau_0 = 0.25$~fm/c.}
\label{f2}
\end{figure} 
%%%%%%%%%%%%%%%%%%%%%%%%%%%%%%%%%%%%%%%%%%%%%%%%%%%%%%%%%

We summarize the results and observations of the analysis in this section. The temperature behaviour of the second-order coefficients of shear tensor and bulk viscous pressure is depicted in Fig.~\ref{f1}. The effect of non-ideal EoS to the second-order coefficients are described in the Eqs.~(\ref{delta pipi})-(\ref{delta PiPi}). Note that the mean field contributions to the coefficients have a visible impact on the temperature regime near transition temperature and vanishes asymptotically with the increase in temperature. Moreover, mean field terms in the analysis are essential to maintain thermodynamic consistency as explained in Ref.~\cite{Jaiswal:2020qmj}. We observe that the non-ideal effects to the shear coefficients are quite significant in the lower temperature regimes (left panel). In the current analysis, the bulk viscous contribution of the system is solely from the medium interactions that are incorporated within the EQPM description. It is important to emphasize that the bulk viscous coefficients vanish in the massless limit with the ideal EoS. Previous studies~\cite{Chandra:2011en, Mitra:2017sjo, Bhadury:2019xdf} have shown that the effective description of the QGP medium with the EQPM breaks the conformal invariance in the massless case. The temperature dependence of second order bulk viscous coefficients is plotted in Fig.~\ref{f1} (right panel). The EQPM results are compared with that from the effective mass quasiparticle model as described in Ref.~\cite{Tinti:2016bav}.

%%%%%%%%%%%%%%%%%%%%%%%%%%%%%%%%%%%%%%%%%%%%%%%%%%%%%%%%%%%
\begin{figure}
\centering
\includegraphics[scale=0.375]{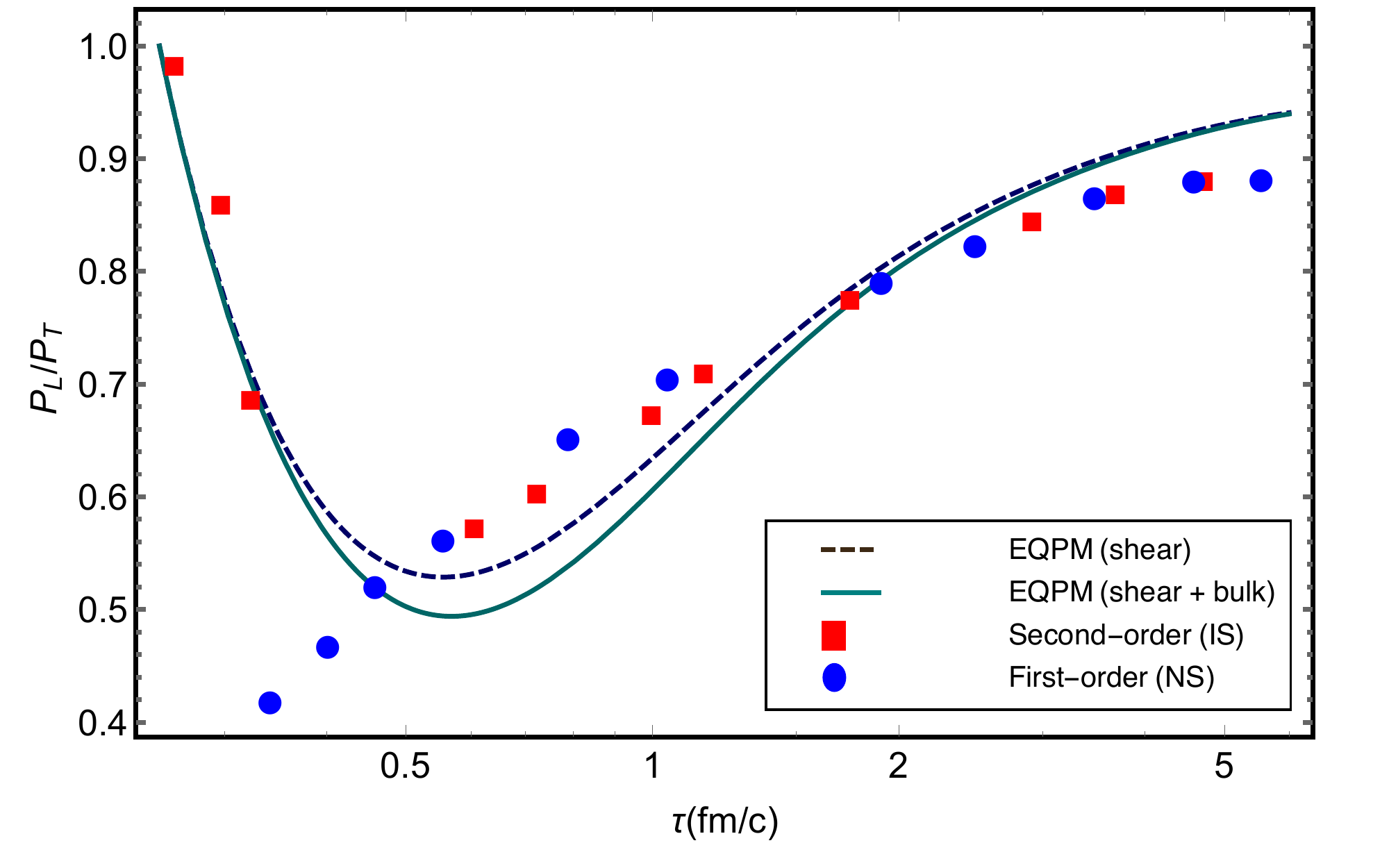}
\includegraphics[scale=0.375]{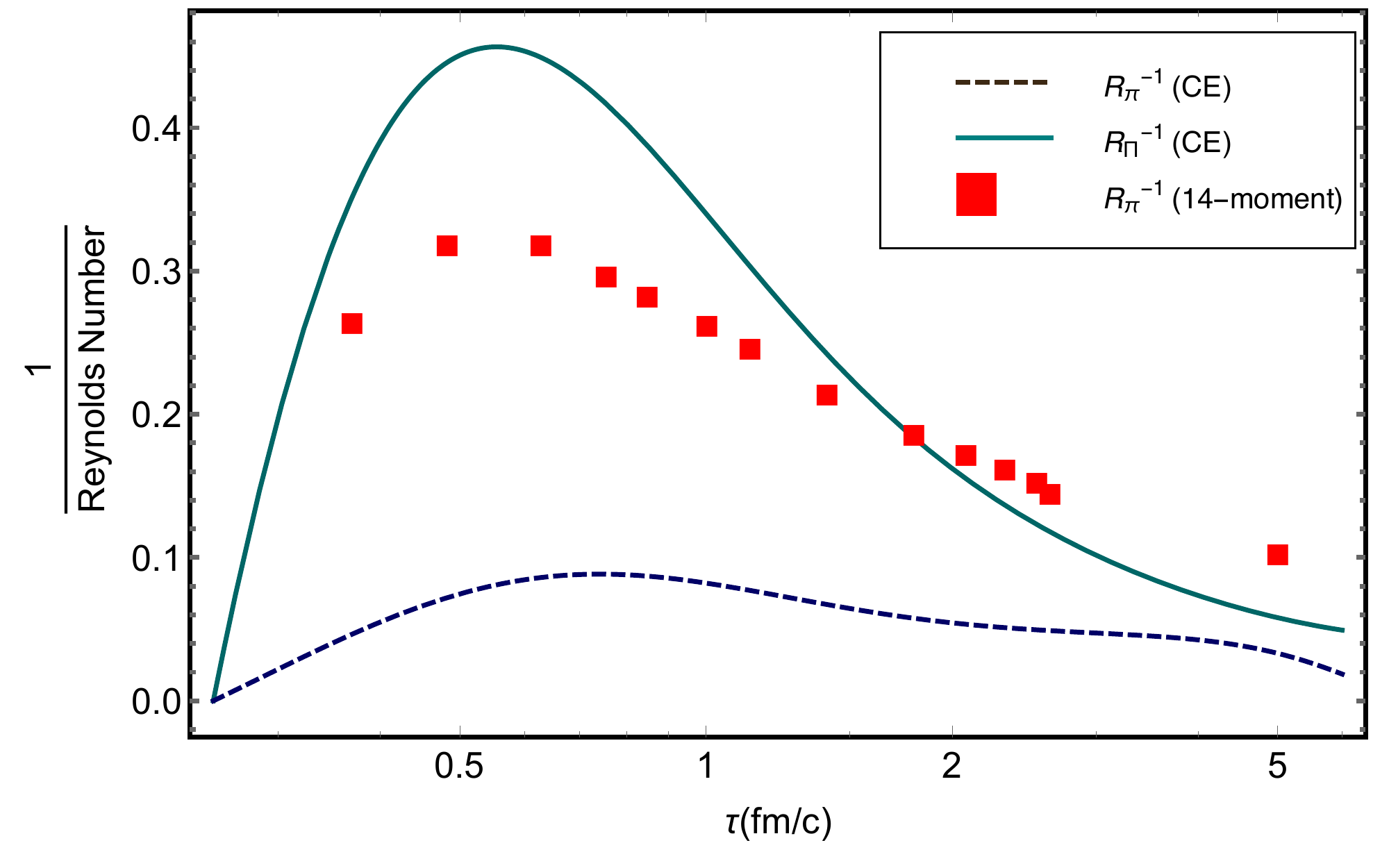}
\caption{ The proper time evolution of pressure anisotropy, $P_L/P_T$ (left panel). The behavior of $P_L/P_T$ is compared with the solutions of IS and NS theory~\cite{Strickland:2014pga}. Proper time dependence of the inverse of Reynolds number associated with shear and bulk viscous pressure (right panel). The evolution of $R^{-1}_\pi$ is compared with that obtained from Grad's 14-moment method~\cite{Mitra:2019jld}. }
\label{f3}
\end{figure} 
%%%%%%%%%%%%%%%%%%%%%%%%%%%%%%%%%%%%%%%%%%%%%%%%%%%%%%%%%%

The effects of the present formulation of second order viscous hydrodynamics of the QGP are quantified for the case of boost invariant longitudinal expansion. We plotted the proper time evolution of temperature in Fig.~\ref{f2} by considering the initial conditions that roughly correspond to those at the LHC. We consider the initial temperature $T_0=600$ MeV at initial proper time $\tau_0=0.25$ fm/c with $\pi(\tau_0)$=0 and $\Pi(\tau_0)$=0. The temperature evolution is studied for the perfect fluid with no viscous effects and for the first and second order viscous hydrodynamics expansion of the medium. While ideal hydrodynamics predicts faster cooling of the medium, the viscous effects slow down temperature evolution. We observe a substantial modification in the evolution of temperature from second order expansion of the medium. The temperature evolution based on second order viscous hydrodynamics drops faster in comparison with the first order hydrodynamic evolution. This observation is consistent with the results of Ref.~\cite{Muronga:2003ta}. The hot QCD medium interactions also affect the evolution of the temperature of the medium. We observe that the thermal medium effect slows down the temperature evolution obtained from the second order hydrodynamic expansion and these effects are more visible in the later stage of evolution. These EoS effects may give corrections to the photon and dilepton spectra, which are sensitive to the temperature evolution of the medium. 

%%%%%%%%%%%%%%%%%%%%%%%%%%%%%%%%%%%%%%%%%%%%%%%%%%%%%%%%%%%%%%%%%%
\begin{figure}
\centering
\includegraphics[scale=0.5]{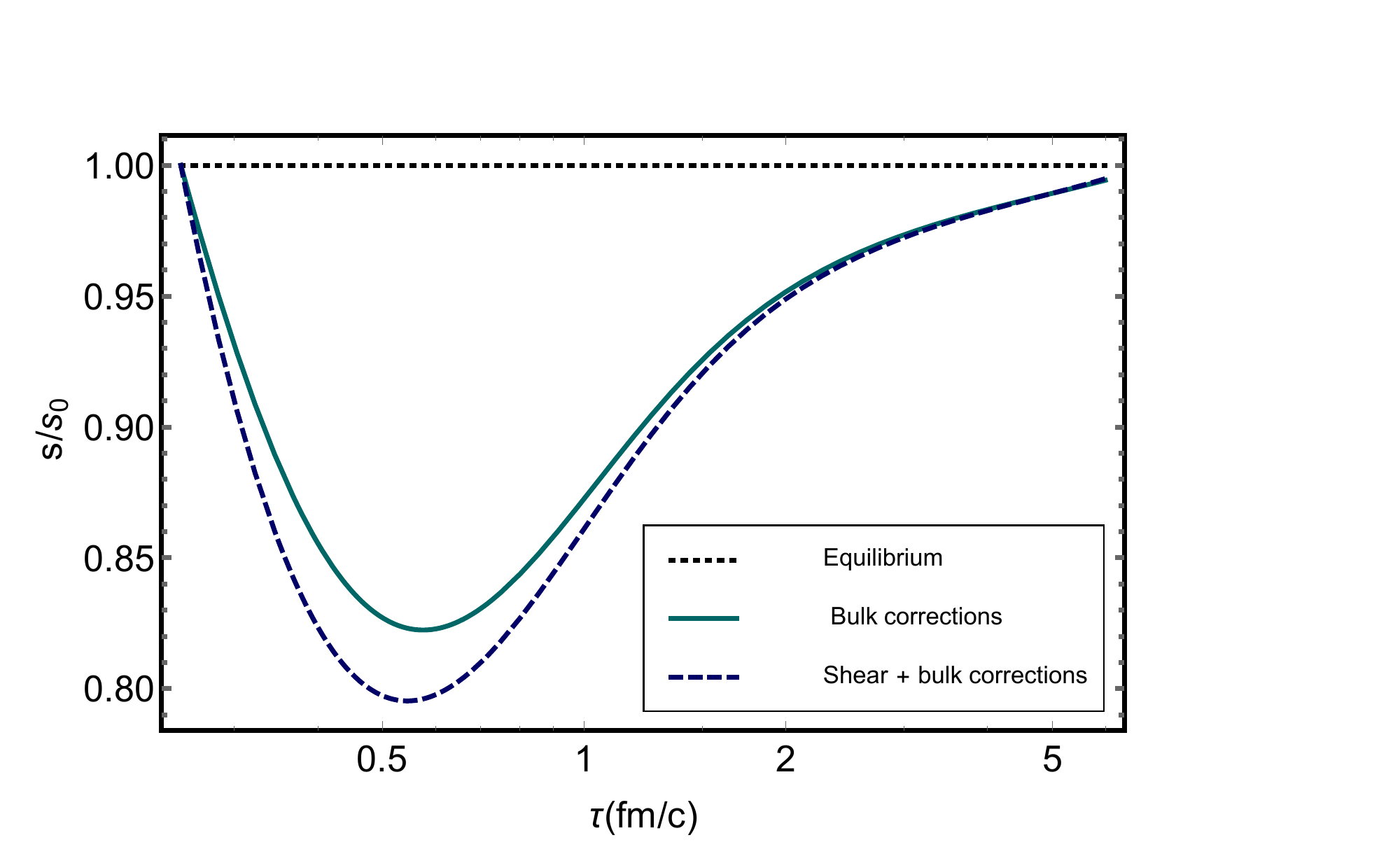}
\caption{ Viscous correction to the proper time evolution of $s/s_0$ with initial temperature $T_0 = 600$ MeV at the initial proper time $\tau_0 = 0.25$~fm/c.}
\label{f4}
\end{figure} 
%%%%%%%%%%%%%%%%%%%%%%%%%%%%%%%%%%%%%%%%%%%%%%%%%%%%%%%%%%%%%%%%

In Fig.~\ref{f3} (left panel), we plotted the proper time dependence of the ratio of longitudinal pressure to the transverse pressure $P_L/P_T$. In a previous study~\cite{Bhadury:2019xdf}, we have realized that the pressure due to shear tensor and bulk viscous pressure is larger than the thermodynamic pressure and hence the ratio $P_L/P_T$ goes to a negative value at the very initial stages of collision with first order EQPM hydrodynamic evolution. In the present analysis, we have studied the time evolution of pressure anisotropy, assuming the initial pressure configuration to be isotropic. We compared the results of pressure anisotropy with the solutions of IS second order viscous hydrodynamical evolution and first order NS theory at $\frac{\eta}{s}=\frac{1}{4\pi}$~\cite{Strickland:2014pga}. We see that the effect of the bulk viscous pressure to the proper time behaviour of $P_L/P_T$ is relatively small in comparison with the pressure from the shear stress tensor. This feature is more evident from the evolution of Reynolds number associated with shear and bulk viscous pressure, $R^{-1}_\pi$ and $R^{-1}_\Pi$, respectively, as plotted in Fig.~\ref{f3} (right panel). We have compared the results of the $R^{-1}_\pi$ with that obtained from the Grad's 14-moment method~\cite{Mitra:2019jld}. Although both the CE expansion and Grad's 14-moment methods involve expanding the distribution function around the equilibrium state, we observe significant differences in the system response followed by the viscous flow. This is reflected in the time evolution of the Reynolds number. 

The proper time evolution of the ratio $s/s_0$ is depicted in Fig.~\ref{f4} considering the initial temperature $T_0=600$ MeV at $\tau_0=0.25$ fm/c obtained using second order CE method. Unlike the non-equilibrium shear part, the bulk viscous correction to the distribution function explicitly depends on the effective fugacity parameter, which, in turn, gives a leading order contribution from first order correction to the entropy current within the EQPM description. We studied the bulk and shear viscous corrections to the entropy current. We observed that the entropy density deviates significantly from the equilibrium value due to the viscous evolution in the intermediate stage. The entropy density attains its equilibrium value at the later stage as the system approaches equilibrium. 

%%%%%%%%%%%%%%%%%%%%%%%%%%%%%%%%%%%%%%%%%%%%%%%%%%%%%%%%%%%%%%%
\section{Summary and Outlook}

We have presented the formulation of the second-order viscous hydrodynamic evolution equations while incorporating the thermal medium effects via realistic QCD equation of state. A quasiparticle model is employed in the analysis to encode the equation of state effects via effective quark and gluon fugacity parameters. The system away from the equilibrium has been modelled by the effective kinetic theory. We have employed CE expansion method to solve the relativistic effective Boltzmann equation with the RTA. We have studied the temperature dependence of the shear and bulk second order transport coefficients within the EQPM and compared the results with other parallel analysis. The hot QCD medium effects induce visible modification to the transport coefficients, especially in the temperature regimes not very far from the transition temperature $T_c$. We also studied the viscous corrections to the entropy four-current using second order CE method. We found non-vanishing entropy flux at second order which appears purely due to the EQPM.

Having obtained the effective description of the second order evolution equations, we then focused on the special case of a boost-invariant longitudinal expansion of the system. The evolution of temperature with proper time is studied in the viscous $1+1-$dimensional Bjorken expanding medium. We observe a faster temperature drop with proper time within the second order hydrodynamic evolution equations in comparison with the first order theory. Notably, the inclusion of thermal medium effects in the second order theory slows down the temperature evolution. The EQPM formulation of second-order hydrodynamic evolution equations has been further employed to study the pressure anisotropy and Reynolds number associated with shear and bulk viscous pressure in the medium. We found that both the equation of state as well as the second-order transport coefficients have a sizable effect in the proper time evolution of pressure anisotropy. We have also compared the proper time dependence of $R^{-1}_\pi$ obtained from the CE expansion and Grad's 14-moment methods. We observed a sizable deviation of the entropy density from its equilibrium value due to the viscous corrections.

Looking forward, the formulation of second-order magnetohydrodynamics with an effective transport theory would be a timely and interesting task to pursue. Apart from this, a thorough understanding of the hydrodynamics description of the medium by employing a more realistic collision kernel, and the phenomenological significance of the second-order transport coefficients in the heavy quark dynamics and dilepton production in the expanding medium are another interesting directions to work in the near future. Moreover, it would be important to formulate a unified framework for effective mass and effective fugacity models such that one can describe the lattice equation of state as well as higher-order susceptibilities within the same framework. We leave these problems for future studies.

%%%%%%%%%%%%%%%%%%%%%%%%%%%%%%%%%%%%%%%%%%%%%%%%%%%%%%%%%%%%%
\section*{ACKNOWLEDGMENTS}
S.B. would like to acknowledge the kind hospitality of IIT Gandhinagar, where part of this work was completed. M.K. would like to acknowledge the hospitality of NISER. We thank Sayanatani Bhattacharya for fruitful discussions on second-order transport coefficients. This research was supported in part by the International Centre for Theoretical Sciences (ICTS) during a visit for participating in the program-QCD matter (Code: ICTS/extremeqandg/2019/04). A.J. was supported in part by the DST-INSPIRE faculty award under Grant No. DST/INSPIRE/04/2017/000038. We are indebted to the people of India for their generous support for the research in basic sciences.
%%%%%%%%%%%%%%%%%%%%%%%%%%%%%%%%%%%%%%%%%%%%%%%%%%%%%%%%%%%%%

\appendix
\section{Thermodynamic integrals in the massless case }\label{A}
The thermodynamic integrals can be expressed in terms of polylogarithm function $\mathrm{PolyLog}~[n, a]$, where $a$ is the argument and $n$ denotes the order, in the massless limit. Note that in the finite quark mass limit, the integrals can be represented with the modified Bessel function of the second kind~\cite{Bhadury:2019xdf}. For the quark case, the integrals take the following forms,
\begin{align}
\Tilde{J}^{(3)}_{q~63} =&~ -\frac{g_qT^5}{210\pi^2} \! \bigg[\! -24\mathrm{PolyLog}~[4,-z_q] + 24\frac{(\delta\omega_q)}{T}\mathrm{PolyLog}~[3,-z_q]-20\frac{(\delta\omega_q)^2}{T^2}\mathrm{PolyLog}~[2,-z_q] \! \bigg],
\end{align}
\begin{align}
\Tilde{J}^{(2)}_{q~42} =&~ \frac{g_qT^4}{5\pi^2}\bigg[-\mathrm{PolyLog}~[3,-z_q]
+\frac{(\delta\omega_q)}{T}\mathrm{PolyLog}~[2,-z_q]+\frac{(\delta\omega_q)^2}{T^2}\mathrm{Log}(1+z_q)\bigg],
\end{align}
\begin{align}
\Tilde{J}^{(1)}_{q~42} =&~ \frac{g_qT^5}{5\pi^2}\bigg[-4\mathrm{PolyLog}~[4,-z_q] + 2\frac{(\delta\omega_q)}{T}\mathrm{PolyLog}~[3,-z_q]-\frac{(\delta\omega_q)^2}{T^2}\mathrm{PolyLog}~[2,-z_q]\bigg],
\end{align}
\begin{align}
\Tilde{J}^{(1)}_{q~31} =&~ -\frac{g_qT^4}{6\pi^2}\bigg[-6\mathrm{PolyLog}~[3,-z_q]
+ 2\frac{(\delta\omega_q)}{T}\mathrm{PolyLog}~[2,-z_q]+\frac{(\delta\omega_q)^2}{T^2}\mathrm{Log}(1+z_q)\bigg],
\end{align}
\begin{align}
\Tilde{J}^{(0)}_{q~21} =&~ -\frac{g_qT^4}{\pi^2}\bigg[-\mathrm{PolyLog}~[3,-z_q]    + \frac{1}{3}\frac{(\delta\omega_q)}{T}\mathrm{PolyLog}~[2,-z_q]+\frac{1}{6}\frac{(\delta\omega_q)^2}{T^2}\mathrm{Log}(1+z_q)\bigg],
\end{align}
\begin{align}
\Tilde{J}^{(3)}_{q~42} =-\frac{g_qT^3}{15\pi^2}\bigg[\mathrm{PolyLog}~[2,-z_q]
+ 2\frac{(\delta\omega_q)}{T}\mathrm{Log}(1+z_q)\bigg],
\end{align}
\begin{align}
\Tilde{J}^{(1)}_{q~21} =-\frac{g_qT^3}{\pi^2}\bigg[-\frac{1}{3}\mathrm{PolyLog}~[2,-z_q]
-\frac{1}{3}\frac{(\delta\omega_q)}{T}\mathrm{Log}(1+z_q)+\frac{1}{2}\frac{(\delta\omega_q)^2}{T^2}\frac{z_q}{1+z_q}\bigg],
\end{align}
\begin{align}
\Tilde{J}^{(0)}_{q~20} =-\frac{g_qT^4}{\pi^2}\bigg[3\mathrm{PolyLog}~[3,-z_q]
+ \frac{(\delta\omega_q)}{T}\mathrm{PolyLog}~[2,-z_q]\bigg],
\end{align}
\begin{align}
\Tilde{J}^{(2)}_{q~21} =-\frac{g_qT^2}{6\pi^2}\bigg[\mathrm{Log}(1+z_q)-3\frac{(\delta\omega_q)}{T}\frac{z_q}{1+z_q}\bigg],
\end{align}
\begin{align}
\Tilde{J}^{(1)}_{q~10} =\frac{g_qT^2}{2\pi^2}\bigg[\mathrm{Log}(1+z_q)-\frac{(\delta\omega_q)}{T}\frac{z_q}{1+z_q}\bigg],
\end{align}
\begin{align}
\Tilde{J}^{(0)}_{q~31} =\frac{g_q T^5}{\pi^2}4\mathrm{PolyLog}~[4,-z_q],
\end{align}
\begin{align}
\Tilde{J}^{(0)}_{q~10} =-\frac{g_q T^3}{\pi^2}\mathrm{PolyLog}~[2,-z_q],
\end{align}
\begin{align}
\Tilde{J}^{(2)}_{q~10} =\frac{g_qT^2}{2\pi^2}\frac{z_q}{1+z_q},
\end{align}
\begin{align}
\Tilde{M}^{(0)}_{q~31} =\frac{g_q T^5}{\pi^2}4\mathrm{PolyLog}~[3,-z_q],
\end{align}
\begin{align}
\Tilde{M}^{(0)}_{q~10} =\frac{g_q T^3}{\pi^2}\mathrm{Log}~(1+z_q),
\end{align}
\begin{align}
\Tilde{M}^{(1)}_{q~10} =\frac{g_q T^2}{2\pi^2}\frac{z_q}{1+z_q},
\end{align}
\begin{align}
\Tilde{M}^{(1)}_{q~21} =-\frac{g_q T^2}{6\pi^2}\frac{z_q}{1+z_q},
\end{align}
\begin{align}
\Tilde{M}^{(1)}_{q~42} =\frac{g_qT^5}{5\pi^2}\bigg[-4\mathrm{PolyLog}~[3,-z_q]
+ 2\frac{(\delta\omega_q)}{T}\mathrm{PolyLog}~[2,-z_q]+\frac{(\delta\omega_q)^2}{T^2}\mathrm{Log}(1+z_q)\bigg],
\end{align}
\begin{align}
\Tilde{M}^{(0)}_{q~30} =\frac{g_qT^5}{\pi^2}\bigg[-12\mathrm{PolyLog}~[3,-z_q]
- 6\frac{(\delta\omega_q)}{T}\mathrm{PolyLog}~[2,-z_q]+\frac{(\delta\omega_q)^2}{T^2}\mathrm{Log}(1+z_q)\bigg],
\end{align}
\begin{align}
\Tilde{M}^{(0)}_{q~21}=-\frac{g_qT^4}{\pi^2}\bigg[-\mathrm{PolyLog}~[2,-z_q]
-\frac{1}{3}\frac{(\delta\omega_q)}{T}\mathrm{Log}(1+z_q)+\frac{1}{6}\frac{(\delta\omega_q)^2}{T^2}\frac{z_q}{1+z_q}\bigg],
\end{align}
\begin{align}
\Tilde{M}^{(0)}_{q~20}=\frac{g_qT^4}{\pi^2}\bigg[-3\mathrm{PolyLog}~[2,-z_q]
+\frac{(\delta\omega_q)}{T}\mathrm{Log}(1+z_q)\bigg],
\end{align}
\begin{align}
\Tilde{L}^{(3)}_{q~63} = -\frac{g_qT^4}{210\pi^2}\bigg[-6\mathrm{PolyLog}~[3,-z_q]+8\frac{(\delta\omega_q)}{T}\mathrm{PolyLog}~[2,-z_q]\bigg],
\end{align}
\begin{align}
\Tilde{L}^{(2)}_{q~42} = \frac{g_qT^3}{5\pi^2}\bigg[-\frac{1}{3}\mathrm{PolyLog}~[2,-z_q]-\frac{1}{2}\frac{(\delta\omega_q)}{T}\mathrm{Log}(1+z_q)\bigg],
\end{align} 
\begin{align}
\Tilde{L}^{(1)}_{q~42} = \frac{g_qT^4}{5\pi^2}\bigg[-\mathrm{PolyLog}~[3,-z_q]+\frac{2}{3}\frac{(\delta\omega_q)}{T}\mathrm{PolyLog}~[2,-z_q]\bigg],
\end{align}
\begin{align}
\Tilde{L}^{(0)}_{q~21}=-\frac{g_qT^3}{\pi^2}\bigg[-\frac{1}{3}\mathrm{PolyLog}~[2,-z_q]-\frac{1}{6}\frac{(\delta\omega_q)}{T}\mathrm{Log}(1+z_q)\bigg],
\end{align} 
\begin{align}
\Tilde{L}^{(1)}_{q~31} = -\frac{g_qT^3}{6\pi^2}\bigg[-2\mathrm{PolyLog}~[2,-z_q]-\frac{(\delta\omega_q)}{T}\mathrm{Log}(1+z_q)\bigg],
\end{align}
\begin{align}   
\Tilde{L}^{(0)}_{q~31} =\frac{g_qT^4}{\pi^2}\mathrm{PolyLog}~[3,-z_q],
\end{align}
\begin{align}   
\Tilde{N}^{(0)}_{q~31} =\frac{g_qT^4}{\pi^2}\mathrm{PolyLog}~[2,-z_q],
\end{align}
\begin{align}
\Tilde{N}^{(1)}_{q~42} = \frac{g_qT^4}{5\pi^2}\bigg[-\mathrm{PolyLog}~[2,-z_q]-\frac{2}{3}\frac{(\delta\omega_q)}{T}\mathrm{Log}(1+z_q)\bigg].
\end{align}
For the gluonic case the thermodynamic integrals take the following forms,
\begin{align}
       \Tilde{J}^{(1)}_{g~42} =\frac{g_gT^5}{5\pi^2}\bigg[4\mathrm{PolyLog}~[4,z_g]
    -2\frac{(\delta\omega_g)}{T}\mathrm{PolyLog}~[3,z_g]+\frac{(\delta\omega_g)^2}{T^2}\mathrm{PolyLog}~[2,z_g]\bigg],
\end{align}
\begin{align}
       \Tilde{J}^{(2)}_{g~42} =\frac{g_gT^4}{5\pi^2}\bigg[\mathrm{PolyLog}~[3,z_g]
    -\frac{(\delta\omega_g)}{T}\mathrm{PolyLog}~[2,z_g]-\frac{(\delta\omega_g)^2}{T^2}\mathrm{Log}(1-z_g)\bigg],
\end{align}
\begin{align}
    \Tilde{J}^{(0)}_{g~21} =-\frac{g_gT^4}{\pi^2}\bigg[\mathrm{PolyLog}~[3,z_g]-\frac{1}{3}\frac{(\delta\omega_g)}{T}\mathrm{PolyLog}~[2,z_g]-\frac{1}{6}\frac{(\delta\omega_g)^2}{T^2}\mathrm{Log}(1-z_g)\bigg]
\end{align}
\begin{align}
       \Tilde{J}^{(3)}_{g~63} =-\frac{g_gT^5}{210\pi^2}\bigg[24\mathrm{PolyLog}~[4,z_g]
    -24\frac{(\delta\omega_g)}{T}\mathrm{PolyLog}~[3,z_g]+20\frac{(\delta\omega_g)^2}{T^2}\mathrm{PolyLog}~[2,z_g]\bigg],
\end{align}
\begin{align}
       \Tilde{J}^{(1)}_{g~31} =-\frac{g_gT^4}{6\pi^2}\bigg[6\mathrm{PolyLog}~[3,z_g]
    -2\frac{(\delta\omega_g)}{T}\mathrm{PolyLog}~[2,z_g]-\frac{(\delta\omega_g)^2}{T^2}\mathrm{Log}(1-z_g)\bigg],
\end{align}
\begin{align}
       \Tilde{J}^{(1)}_{g~21} =-\frac{g_gT^3}{\pi^2}\bigg[\frac{1}{3}\mathrm{PolyLog}~[2,z_g]
    +\frac{1}{3}\frac{(\delta\omega_g)}{T}\mathrm{Log}(1-z_g)+\frac{1}{2}\frac{(\delta\omega_g)^2}{T^2}\frac{z_g}{1-z_g}\bigg],
\end{align}
\begin{align}
       \Tilde{J}^{(3)}_{g~42} =\frac{g_gT^3}{15\pi^2}\bigg[\mathrm{PolyLog}~[2,z_g]
    +2\frac{(\delta\omega_g)}{T}\mathrm{Log}(1-z_g)\bigg],
\end{align}
\begin{align}
       \Tilde{J}^{(0)}_{g~20} =\frac{g_gT^4}{\pi^2}\bigg[3\mathrm{PolyLog}~[3,z_g]
    +\frac{(\delta\omega_g)}{T}\mathrm{PolyLog}~[2,z_g]\bigg],
\end{align}
\begin{align}
    \Tilde{J}^{(2)}_{g~21} =-\frac{g_gT^2}{6\pi^2}\bigg[-\mathrm{Log}(1-z_g)-3\frac{(\delta\omega_g)}{T}\frac{z_g}{1-z_g}\bigg],
\end{align}
\begin{align}
    \Tilde{J}^{(1)}_{g~10} =\frac{g_gT^2}{2\pi^2}\bigg[-\mathrm{Log}(1-z_g)-\frac{(\delta\omega_g)}{T}\frac{z_g}{1-z_g}\bigg],
\end{align}
\begin{align}
    \Tilde{J}^{(0)}_{g~31} =-\frac{g_g~T^5}{\pi^2}~4\mathrm{PolyLog}~[4,z_g],
\end{align}
\begin{align}
    \Tilde{J}^{(0)}_{g~10} =\frac{g_g~T^3}{\pi^2}~\mathrm{PolyLog}~[2,z_g],
\end{align}
\begin{align}
    \Tilde{J}^{(2)}_{g~10} =\frac{g_gT^2}{2\pi^2}\frac{z_g}{1-z_g},
\end{align}
\begin{align}
    \Tilde{M}^{(0)}_{g~31} =-\frac{g_gT^5}{\pi^2}~4\mathrm{PolyLog}~[3,z_g],
\end{align}
\begin{align}
    \Tilde{M}^{(0)}_{g~10} =-\frac{g_gT^3}{\pi^2}~\mathrm{Log}~(1-z_g),
\end{align}
\begin{align}
    \Tilde{M}^{(1)}_{g~10} =\frac{g_gT^2}{2\pi^2}~\frac{z_g}{1-z_g},
\end{align}
\begin{align}
    \Tilde{M}^{(1)}_{g~21} =-\frac{g_gT^2}{6\pi^2}~\frac{z_g}{1-z_g},
\end{align}
\begin{align}
       \Tilde{M}^{(1)}_{g~42} =\frac{g_gT^5}{5\pi^2}\bigg[4\mathrm{PolyLog}~[3,z_g]
    -2\frac{(\delta\omega_g)}{T}\mathrm{PolyLog}~[2,z_g]-\frac{(\delta\omega_g)^2}{T^2}\mathrm{Log}(1-z_g)\bigg],
\end{align}
\begin{align}
       \Tilde{M}^{(0)}_{g~30} =\frac{g_gT^5}{\pi^2}\bigg[12\mathrm{PolyLog}~[3,z_g]
    +6\frac{(\delta\omega_g)}{T}\mathrm{PolyLog}~[2,z_g]-\frac{(\delta\omega_g)^2}{T^2}\mathrm{Log}(1-z_g)\bigg],
\end{align}
\begin{align}
    \Tilde{M}^{(0)}_{g~21} =-\frac{g_gT^4}{\pi^2}\bigg[\mathrm{PolyLog}~[2,z_g]+\frac{1}{3}\frac{(\delta\omega_g)}{T}\mathrm{Log}(1-z_g)+\frac{1}{6}\frac{(\delta\omega_g)^2}{T^2}\frac{z_g}{1-z_g}\bigg],
\end{align}
\begin{align}
    \Tilde{M}^{(0)}_{g~20} =\frac{g_gT^4}{\pi^2}\bigg[3\mathrm{PolyLog}~[2,z_g]-\frac{(\delta\omega_g)}{T}\mathrm{Log}(1-z_g)\bigg],
\end{align}
\begin{align}
    \Tilde{L}^{(3)}_{g~63} =-\frac{g_gT^4}{210\pi^2}\bigg[6\mathrm{PolyLog}~[3,z_g]-8\frac{(\delta\omega_g)}{T}\mathrm{PolyLog}~[2,z_g]\bigg],
\end{align}
\begin{align}
    \Tilde{L}^{(2)}_{g~42} =\frac{g_gT^3}{5\pi^2}\bigg[\frac{1}{3}\mathrm{PolyLog}~[2,z_g]+\frac{1}{2}\frac{(\delta\omega_g)}{T}\mathrm{Log}(1-z_g)\bigg],
\end{align}
\begin{align}
    \Tilde{L}^{(1)}_{g~42} =\frac{g_gT^4}{5\pi^2}\bigg[\mathrm{PolyLog}~[3,z_g]-\frac{2}{3}\frac{(\delta\omega_g)}{T}\mathrm{PolyLog}~[2,z_g]\bigg],
\end{align}
\begin{align}
    \Tilde{L}^{(0)}_{g~21} =-\frac{g_gT^3}{\pi^2}\bigg[\frac{1}{3}\mathrm{PolyLog}~[2,z_g]+\frac{1}{6}\frac{(\delta\omega_g)}{T}\mathrm{Log}(1-z_g)]\bigg]
\end{align}
\begin{align}
    \Tilde{L}^{(0)}_{g~31} =-\frac{g_gT^4}{\pi^2}\mathrm{PolyLog}~[3,z_g],
\end{align} 
\begin{align}
    \Tilde{N}^{(1)}_{g~42} =\frac{g_gT^4}{5\pi^2}\bigg[\mathrm{PolyLog}~[2,z_g]+\frac{2}{3}\frac{(\delta\omega_g)}{T}\mathrm{Log}(1-z_g)\bigg],
\end{align}
\begin{align}
    \Tilde{L}^{(1)}_{g~31} =-\frac{g_gT^3}{6\pi^2}\bigg[2\mathrm{PolyLog}~[2,z_g]+\frac{(\delta\omega_g)}{T}\mathrm{Log}(1-z_g)\bigg],
\end{align}
\begin{align}
    \Tilde{N}^{(0)}_{g~31} =-\frac{g_gT^4}{\pi^2}\mathrm{PolyLog}~[2,z_g].
\end{align} 
%
%%%%%%%%%%%%%%%%%%%%%%%%%%%%%%%%%%%%%%%%%%%%%%%%%%%%%%%%%%
%
\section{Coefficients associated to entropy current} \label{B}
The $C_i$ coefficients of Eqs.~\eqref{phi 1} and \eqref{phi 2} are given by,
%1
\begin{align}
   {C}_{1k} &= \frac{\beta}{2\, \beta_\pi (u\cdot \Tilde{p}_k)},\label{C1}
\end{align}
%2
\begin{align}
    %2
    {C}_{2k} &= \beta\left[\frac{\chi_{3k}}{\beta_\Pi} - \frac{\chi_{1k} (u\cdot \Tilde{p}_k)}{\beta_\Pi}\right],
\end{align}
%3
\begin{align}
    {C}_{3k} &= \beta\, \tau_R\left[\frac{1}{ \beta_\pi} + \frac{1}{\varepsilon + P}\right] + \frac{3\, \beta\, \tau_R (\delta\omega_k)}{2\, \beta_\pi (u\cdot\Tilde{p}_k)},
\end{align}
%4
\begin{align}
    {C}_{4k} &= \frac{\beta\, \tau_R}{\beta_\pi (u\cdot\Tilde{p}_k)^2} - \frac{\beta^2\, \tau_R\, \chi_{\pi k}}{2\, \beta_\pi^2 (u\cdot\Tilde{p}_k)^2} - \frac{\beta^2\, \tau_R}{2\, \beta_\pi (u\cdot\Tilde{p}_k)^3} \frac{\partial ( \delta\omega_k)}{\partial\beta},
\end{align}
%5
\begin{align}
    {C}_{5k} &= - \frac{\beta\, \tau_R}{\varepsilon + P},
\end{align}
%6
\begin{align}
    {C}_{6k} &= - \frac{\beta \tau_R}{2 \beta_\pi (u\cdot\Tilde{p}_k)^2},
\end{align}
%7
\begin{align}
    {C}_{7k} &= \frac{\beta}{2\, \beta_\pi} \left[ \frac{\lambda_{\Pi\pi}\, \chi_{1k} (u \cdot \Tilde{p}_k)}{\beta_\Pi} - \frac{(u\cdot \Tilde{p}_k) c_s^2}{\varepsilon + P} + \frac{\tau_{\pi\pi} \chi_{2k}}{2\, \beta_\pi (u\cdot \Tilde{p}_k)} - \frac{(u\cdot \Tilde{p}_k)}{\varepsilon + P}  + \chi_{4k} - \frac{\chi_{3k}\, \lambda_{\Pi\pi}}{\beta_\Pi} \!\right],
\end{align}
%8
\begin{align}
    {C}_{8k} &= \frac{\beta\, \tau_{\pi\pi}}{4\, \beta_\pi^2 (u\cdot \Tilde{p}_k)} - \frac{\beta\, (\delta\omega_k)}{2\, \beta_\pi^2 (u\cdot\Tilde{p}_k)^2},
\end{align}
%9
\begin{align}
    {C}_{9k} &= \frac{\beta}{4\, \beta_\pi^2 (u\cdot \Tilde{p}_k)^3} \Big[1 + \beta (u\cdot\Tilde{p}_k)(\bar{f}^0_k - a f^0_k)\Big],
\end{align}
%10
\begin{align}
    {C}_{10k} &= - \frac{\beta\, \tau_R}{\beta_\pi (u\cdot \Tilde{p}_k)} - \frac{\beta\, \tau_R (\delta\omega_k)}{\beta_\pi (u\cdot\Tilde{p}_k)^2},
\end{align}
%11
\begin{align}
    {C}_{11k} &= - \frac{\beta}{\beta_\Pi}\frac{\delta_{\pi\pi}}{2\, \beta_\pi (u\cdot \Tilde{p}_k)} - \frac{c_s^2}{2\, \beta_\pi (u\cdot\Tilde{p}_k)} - \frac{(\delta\omega_k)}{3\,\beta_\pi (u\cdot\Tilde{p}_k)^2} \!+\! \frac{\beta c_s^2 \chi_{\pi k}}{2\, \beta_\pi^2 (u\cdot\Tilde{p}_k)} \!+\! \frac{1}{6\, \beta_\pi (u\cdot\Tilde{p}_k)^3} \Tilde{p}^\mu_k\, \Tilde{p}_k^\phi \Delta_{\mu\phi} \nonumber\\
    &~~~\!+\! \frac{\beta\, c_s^2}{2\, \beta_\pi (u\cdot\Tilde{p}_k)^2} \frac{\partial( \delta\omega_k)}{\partial\beta} + \frac{\beta}{6 \beta_\pi (u\cdot \Tilde{p}_k)^2} \!\Bigg[\!3 (u\cdot \Tilde{p}_k)^2 c_s^2 + \Tilde{p}^\mu_k \Tilde{p}^\phi_k \Delta_{\mu\phi}\Bigg] \! \left(\bar{f}_k^0 - a f_k^0 \right)\nonumber\\
    &~~~- \frac{\beta}{2 \beta_{\pi}\left(u \cdot \tilde{p}_{k}\right)} \!\Bigg[\!\frac{\partial\left(\delta \omega_{k}\right)}{\partial \beta}  \! \beta c_s^2 + \left(\delta \omega_{k}\right) \Bigg]  \!\! \left(\bar{f}_{k}^{0} - a f_{k}^{0}\right) - \frac{\beta\, \lambda_{\pi\Pi}}{4\, \beta_\pi^2 (u\cdot \Tilde{p}_k)} + \frac{\chi_{6k}\, \beta}{2\, \beta_\pi\, \beta_\Pi} + \frac{\chi_{1k}\, \beta}{2\, \beta_\pi\, \beta_\Pi (u\cdot\Tilde{p}_k)}\nonumber\\
    &- \frac{\beta^2}{2\, \beta_\pi\, \beta_\Pi (u\cdot\Tilde{p}_k)} \Big(\chi_{1k} (u\cdot \Tilde{p}_k) - \chi_{3k}\Big) (\bar{f}_k^0 - a f_k^0),
\end{align}
%12
\begin{align}
    {C}_{12k} &= - \frac{\beta\, \tau_R}{\varepsilon + P} - \frac{\chi_{1k}\, \beta\, \tau_R}{\beta_\Pi} - \frac{\beta\, \tau_R}{\beta_\Pi}\! \Bigg[\! \beta \chi_{5k} \!- \chi_{6k} (u\cdot\Tilde{p}_k) -\! \frac{\beta \chi_{8k}}{(u\cdot\tilde{p}_k)} \Bigg] + \frac{\chi_{1k}\, \beta\, \tau_R}{\beta_\Pi (u\cdot\Tilde{p}_k)} \!\! \nonumber\\
    &~~~\times\Bigg[\!(u \cdot \Tilde{p}_k) \!- \!\beta\! \frac{\partial(\delta\omega_k)}{\partial\beta}  \!\Bigg]+ \Bigg[\frac{\beta^2\, \tau_R\, \chi_\Pi (\chi_{1k} (u\cdot \Tilde{p}_k) - \chi_{3k})}{\beta_\Pi^2  (u\cdot\Tilde{p}_k)}\Bigg] + \frac{\beta\, \tau_R}{\beta_\Pi} \Bigg[\chi_{1k} - \frac{\chi_{3k} }{(u\cdot\tilde{p}_k)}\Bigg],
\end{align}
%13
\begin{align}
    {C}_{13k} &= \frac{\beta\, \tau_R}{\beta_\Pi} \Bigg[\chi_{1k} - \frac{\chi_{3k} }{(u\cdot\tilde{p}_k)} \Bigg] + \frac{\beta\, \tau_R}{\varepsilon + P},
    \end{align}
%14
    \begin{align}
    {C}_{14k} &= - \frac{\beta}{\beta_\Pi} \left[\frac{(u\cdot \Tilde{p}_k) c_s^2}{\varepsilon + P} - \frac{\delta_{\Pi\Pi}\, \chi_{1k} (u \cdot \Tilde{p}_k)}{\beta_\Pi} + \frac{\chi_{3k}\, \delta_{\Pi\Pi}}{\beta_\Pi} - \chi_{4k}\right] - \frac{\chi_{1k}\, \beta}{\beta_\Pi^2} (u\cdot\Tilde{p}_k)\, c_s^2 \nonumber\\
    &~~~- \frac{\beta}{3 \beta_\Pi^2} \!\Bigg[3 \beta c_s^2 \big(\chi_{5k} (u \cdot \Tilde{p}_k) - \chi_{8k}\big)\!+\! \chi_{6k} \Tilde{p}^\mu_k \Tilde{p}_k^\phi \Delta_{\mu \phi}\Bigg] - \frac{\chi_{1k} \beta}{3 \beta_\Pi^2 (u\cdot\Tilde{p}_k)} \!\! \Bigg[ \Tilde{p}^\mu_k\Tilde{p}^\phi_k \Delta_{\mu \phi} + \beta (u\cdot\Tilde{p}_k)\!\! \frac{\partial(\delta\omega_k)}{\partial\beta}\! c_s^2\Bigg]\nonumber\\
    &~~~+ \Bigg[\frac{\beta^2\, \chi_{\Pi k} \big(\chi_{1k} (u\cdot \Tilde{p}_k) - \chi_{3k}\big)}{\beta_\Pi^3}\Bigg] c_s^2 + \frac{\beta^2}{\beta_\Pi^2 (u\cdot\Tilde{p}_k)} \Big( \chi_{1k} (u\cdot \Tilde{p}_k) - \chi_{3k}\Big) \Bigg[(u\cdot \Tilde{p}_k)^2 c_s^2 + \Tilde{p}^\mu_k \Tilde{p}^\phi_k \frac{\Delta_{\mu\phi} }{3}\Bigg] \nonumber\\
    &~~~\times(\bar{f}_k^0- a f_k^0) + \frac{\beta}{\beta_\Pi^2} \Bigg[\frac{\partial\left(\delta \omega_{k}\right)}{\partial \beta} \beta c_{s}^{2} + \left(\delta \omega_{k}\right)\Bigg] \Bigg[\Big(\chi_{7 k}\left(u \cdot \tilde{p}_{k}\right) + \chi_{1k}\Big) - \beta \Big(\chi_{1k}\left(u \cdot \tilde{p}_{k} \right) - \chi_{3k}\Big) \nonumber\\
    &~~~\times\left(\bar{f}_{k}^{0} - a f_{k}^{0}\right)\!\Bigg].\label{C15}
\end{align}
In the Eqs.~\eqref{C1}-\eqref{C15} we made use of some $\chi_{i}$-coefficients. They are obtained as, 
\begin{align}
    \chi_{1\,k} &= c_s^2 + \frac{1}{3\, (u\cdot \Tilde{p}_k)^2} \Big[m^2 - (u\cdot \Tilde{p}_k)^2\Big] + \frac{(\delta\omega_k)}{3\, (u\cdot \Tilde{p}_k)^2}\Big[ 2 (u\cdot \Tilde{p}_k) - (\delta\omega_k)\Big], \label{chi 1}
\end{align}
\begin{align}
    \chi_{2\,k} &= - \frac{1}{3} \Big[\big(m^2 - (u\cdot \Tilde{p}_k)^2\big) + (\delta\omega_k) \big(2 (u\cdot \Tilde{p}_k) - (\delta\omega_k) \big)\Big], \label{chi 2}
\end{align}
\begin{align}
    \chi_{3\,k} &= \beta c_s^2 \frac{\partial\delta \omega_k}{\partial \beta} + (\delta \omega_k) \label{chi 3}
\end{align}
\begin{align}
    \chi_{4\,k} &= \frac{\beta c_s^2}{\varepsilon + P} \frac{\partial(\delta \omega_k)}{\partial \beta},\label{chi 4}
\end{align}
\begin{align}
    \chi_{5k} &= \frac{\partial c_s^2}{\partial\beta} + \frac{\partial\delta\omega_k}{\partial\beta} \bigg[ \frac{2}{3\, (u\cdot\Tilde{p}_k)} -\frac{2\, m^2}{3\, (u\cdot \Tilde{p}_k)^3} - \frac{4\, (\delta\omega_k)}{3 (u\cdot \Tilde{p}_k)^2} + \frac{(\delta\omega_k)^2}{3 (u\cdot\Tilde{p}_k)^3}\bigg],\label{chi 5}
\end{align}
\begin{align}
    \chi_{6k} &= - \frac{2\, m^2}{3\, (u\cdot \Tilde{p}_k)^3} - \frac{2\, (\delta\omega_k)}{3 (u\cdot\Tilde{p}_k)^2} + \frac{(\delta\omega_k)^2}{3 (u\cdot \Tilde{p}_k)^3},\label{chi 6}
\end{align}
\begin{align}
    \chi_{7k} &= - \frac{2}{3\, (u\cdot \Tilde{p}_k)^3} \bigg[m^2 + (\delta\omega_k) (u\cdot \Tilde{p}_k) - (\delta\omega_k)^2 \bigg],\label{chi 7}
\end{align}
\begin{align}
    \chi_{8k} &= \frac{\partial(\delta \omega_k)}{\partial \beta} \bigg[(1 + c_s^2) + \beta\frac{\partial c_s^2}{\partial\beta} \bigg] + \beta c_s^2 \frac{\partial^2(\delta \omega_k)}{\partial \beta^2},
\end{align}
\begin{align}
    \chi_{\pi\,k} &= \sum_{k=q,g} \bigg[ -  \frac{\partial(\delta\omega_k)}{\partial\beta} \left(2\, \beta  \Tilde{J}^{(2)}_{k,\, 42} - \Tilde{L}_{k\,42}^{(1)} + 2\, \beta (\delta\omega_k) \Tilde{L}^{(2)}_{k,\, 42} \right) + \left(\Tilde{J}_{k\,42}^{(1)} - \beta\, \Tilde{M}^{(1)}_{k,\, 52}\right) \nonumber\\
    &~~~+ (\delta\omega_k) \left(\Tilde{L}_{k\,42}^{(1)} - \beta (\delta\omega_k) \Tilde{N}^{(1)}_{k,\, 52}\right)\bigg],\label{chi pi}
\end{align}
\begin{align}
    \chi_{\Pi\,k} &= \beta\! \left(\frac{\partial c_s^2}{\partial\beta}\right) \!\! \bigg[\Tilde{J}_{k~31}^{(0)} + (\delta\omega_k) \Tilde{L}_{k~31}^{(0)}\bigg] + c_s^2 \bigg[\!\Tilde{J}_{k~31}^{(0)} + \delta\omega_k \Tilde{L}_{k~31}^{(0)}\!\bigg] + \frac{5}{3} \!\bigg[\! \Tilde{J}_{k~42}^{(1)} + (\delta\omega_k) \Tilde{L}_{k~42}^{(1)} \!\bigg] \nonumber\\
    &~~~- (\delta\omega_k) \Tilde{J}_{k~21}^{(0)} - \beta \! \bigg[\! c_s^2 \Tilde{M}^{(0)}_{k,\, 41} + \frac{5}{3} \Tilde{M}^{(1)}_{k,\, 52} \!\bigg]- \beta (\delta\omega_k) \! \bigg[\! c_s^2 \Tilde{N}^{(0)}_{k,\, 41} + \frac{5}{3} \Tilde{N}^{(1)}_{k,\, 52} - \Tilde{M}^{(0)}_{k,\, 31} \!\bigg]\nonumber\\
    &~~~ + \beta \!\frac{\partial (\delta\omega_k)}{\partial\beta}~\bigg[\! c_s^2 \Tilde{L}_{k~31}^{(0)} - \Tilde{J}_{k~21}^{(0)} - \frac{10}{3} \left(\Tilde{J}_{k~42}^{(2)} - \Tilde{L}_{k~42}^{(1)}\right) + (\delta\omega_k)\left(\Tilde{J}_{k~21}^{(1)} - 2\, \Tilde{L}_{k~42}^{(2)} \right)\!\bigg],\label{chi Pi k}
\end{align}
The $\Lambda$ coefficients in the Eqs.~(\ref{entropy density}) and~(\ref{entropy flux}) are derived as,
\begin{align}
%3
\Lambda_1 =&~ \frac{\beta \tau_R}{\beta_\pi}\sum_k \ln{\left(z_{1k}\right)} \! \left[\left(1 + \frac{\beta_\pi}{(\varepsilon + P)}\right) \Tilde{J}^{(0)}_{k,\, 21} + \frac{3 (\delta\omega_k)}{2} \Tilde{J}^{(1)}_{k,\, 21}
%4
2\, \Tilde{J}^{(2)}_{k,\, 42} - \frac{\beta \chi_\pi}{\beta_\pi} \Tilde{J}^{(2)}_{k,\, 42} - \beta \left(\frac{\partial ( \delta\omega_k) }{\partial\beta} \right) \Tilde{J}^{(3)}_{k,\,42}\right],\label{Lambda 1}
\end{align}
\begin{align}
\Lambda_2 =&~ - \frac{\beta \tau_R}{(\varepsilon + P)} \sum_k \ln{\left(z_{1k}\right)} \left[\Tilde{J}^{(0)}_{k,\, 21}
%6
+ \Tilde{J}^{(2)}_{k,\, 42}\right],\label{Lambda 2}
\end{align}
\begin{align}
\Lambda_3 =&~ \frac{\beta \tau_R}{3\, \beta_\Pi} \sum_k \! \ln{\left(z_{1k}\right)} \! \bigg[ \!\!-\! \frac{3\, \beta_\Pi}{(\varepsilon + P)} \Tilde{J}^{(0)}_{k,\, 21} + 3 \! \left(\! \beta \chi_{8k} \!-\! \frac{\beta \chi_{\Pi} \chi_{3k}}{\beta_\Pi} \!-\! \chi_{3k} \! \right) \! \Tilde{J}^{(1)}_{k,\, 21}\nonumber\\
&- \left(\!\left(2\, m^2 - (\delta\omega_k)^2\right) \Tilde{J}^{(3)}_{k,\, 31} + 2\, (\delta\omega_k) \Tilde{J}^{(2)}_{k,\, 31} \right) - 3 \beta \left(\frac{\partial c_s^2}{\partial\beta}\right) \Tilde{J}^{(0)}_{k,\, 21} \nonumber\\
&+ \left(\frac{\beta\, \chi_{\Pi }}{\beta_\Pi} + 1 \right) \left(\left(3 c_s^2 - 1\right)\Tilde{J}^{(0)}_{k,\, 21} + \left(m^2 -(\delta\omega_k)^2\right) \Tilde{J}^{(2)}_{k,\, 21} + 2 (\delta\omega_k) \Tilde{J}^{(1)}_{k,\, 21}\right) \nonumber\\
&- \left(\! \frac{\partial(\delta\omega_k)}{\partial\beta} \!\right) \!\! \left( \! \left(3 c_s^2 - 1\right) \! \Tilde{J}^{(1)}_{k,\, 21} \!+\! \left(m^2 - (\delta\omega_k)^2\right)\! \Tilde{J}^{(3)}_{k,\, 21} \!+\! 2 (\delta\omega_k) \Tilde{J}^{(2)}_{k,\, 21} \! \right) \nonumber\\
&- \beta \! \left(\!\frac{\partial\delta\omega_k}{\partial\beta}\right) \!\! \left(\! \Tilde{J}^{(1)}_{k,\, 21} - \left(2 m^2 - (\delta\omega_k)^2\right) \Tilde{J}^{(3)}_{k,\, 21} - 4 (\delta\omega_k)\Tilde{J}^{(2)}_{k,\, 21} \!\right)\!\!\bigg],\label{Lambda 3}
\end{align}
\begin{align}
\Lambda_4 =&~ \frac{\beta\, \tau_R}{3\, \beta_\Pi}\sum_k \ln{\left(z_{1k}\right)} \! \left[\left(3 c_s^2 - 1 + \frac{3\, \beta_\Pi}{(\varepsilon + P)}\right) \Tilde{J}^{(0)}_{k,\, 21} + \left(2(\delta\omega_k) - 3\, \chi_{3k}\right) \Tilde{J}^{(1)}_{k,\, 21} + \left(m^2- (\delta\omega_k)^2\right) \Tilde{J}^{(2)}_{k,\, 21}\right],\label{Lambda 4}
\end{align}
\begin{align}
\Lambda_\Pi &=~ \frac{\beta}{3\, \beta_\Pi} \sum_k \ln{\left(z_{1k}\right)} \left[ 3\, \chi_{3k}\, \Tilde{J}^{(0)}_{k,\, 10} - \left(3\, c_s^2 - 1\right) \Tilde{J}^{(0)}_{k,\, 20} + \left( m^2 - (\delta\omega_k)^2 \right) \Tilde{J}^{(2)}_{k,\, 20} + 2 (\delta\omega_k) \Tilde{J}^{(1)}_{k,\, 20} \right],\label{Lambda Pi}\nonumber
\end{align}
\begin{align}
\Lambda_{\pi\pi} =&~ \frac{\beta}{2\, \beta_\pi^2} \sum_k \ln{\left(z_{1k}\right)} \Bigg[ \frac{\beta_\pi}{(\varepsilon + P)} \Tilde{J}^{(0)}_{k,\, 20}
+ \Tilde{J}^{(2)}_{k,\, 41} 
+ \beta_\pi \left(\chi_{4k} - \frac{\chi_{3k}\, \lambda_{\Pi\pi}}{\beta_\Pi} \right) \Tilde{J}^{(0)}_{k,\, 10} - \frac{\beta_\pi (1 + c_s^2)}{(\varepsilon + P)} \Tilde{J}^{(0)}_{k,\, 20}\nonumber\\
&+ \frac{\beta_\pi \lambda_{\Pi\pi}}{3\, \beta_\Pi} \left(\left(3 c_s^2 - 1\right) \Tilde{J}^{(0)}_{k,\, 20} + 2 (\delta\omega_k) \Tilde{J}^{(1)}_{k,\, 20} + \left(m^2 - (\delta\omega_k)^2\right) \Tilde{J}^{(2)}_{k,\, 20}\right)\nonumber\\
&- \frac{\tau_{\pi\pi}}{6}\left(m^2 \Tilde{J}^{(1)}_{k,\, 10} -  \Tilde{J}^{(0)}_{k,\, 20} + 2 (\delta\omega_k) \Tilde{J}^{(0)}_{k,\, 10} - (\delta\omega_k)^2 \Tilde{J}^{(1)}_{k,\, 10}\right)\nonumber\\
&+ \frac{1}{2} \left(\tau_{\pi\pi} \Tilde{J}^{(1)}_{k,\, 31} - 2 (\delta\omega_k)  \Tilde{J}^{(2)}_{k,\, 31}\right)
+ 2 \left( \Tilde{J}^{(2)}_{k,\, 42} + \beta \Tilde{M}^{(1)}_{k,\, 42} \right)\Bigg] 
- \frac{\beta^2}{4\, \beta_\pi^2} \Tilde{J}^{(1)}_{k,\,42}, 
\end{align}

\begin{align}
\Lambda_{\Pi\Pi} =&~ \frac{\beta}{\beta_\Pi^2} \sum_k \ln{\left(z_{1k}\right)} \Bigg[-\frac{\beta_\Pi\, c_s^2}{(\varepsilon + P)} \Tilde{J}^{(0)}_{k,\, 20} 
+ \frac{\delta_{\Pi\Pi}}{3} \! \left( \left(3 c_s^2 - 1\right) \Tilde{J}^{(0)}_{k,\, 20} + \left(m^2 - (\delta\omega_k)^2\right)  \Tilde{J}^{(2)}_{k,\, 20} + 2 (\delta\omega_k) \Tilde{J}^{(1)}_{k,\, 20}\right)\nonumber\\
&- \delta_{\Pi\Pi}\chi_{3k}\, \Tilde{J}^{(0)}_{k,\,10}
+ (\beta_\Pi\, \chi_{4k}) \Tilde{J}^{(0)}_{k,\, 10}
- \frac{c_s^2}{3} \left[\left(3 c_s^2 - 1\right) \Tilde{J}^{(0)}_{k,\, 20} + \left(m^2 - (\delta\omega_k)^2\right) \Tilde{J}^{(2)}_{k,\, 20} + 2 (\delta\omega_k) \Tilde{J}^{(1)}_{k,\, 20}\right] \nonumber\\
&- \frac{\beta\, c_s^2}{3} \bigg( 3 \left(\frac{\partial c_s^2}{\partial\beta}\right) \Tilde{J}^{(0)}_{k,\, 20} + \left(\frac{\partial (\delta\omega_k)}{\partial\beta}\right) \Big( 2 \Tilde{J}^{(0)}_{k,\, 10} - \left(2\, m^2 - (\delta\omega_k)^2\right) \Tilde{J}^{(2)}_{k,\, 10} - 4\, (\delta\omega_k) \Tilde{J}^{(1)}_{k,\, 10}\Big) \bigg)\nonumber\\
&+ \frac{\chi_{8k}}{3} \Tilde{J}^{(0)}_{k,\, 10}
+ \frac{1}{3} \! \left(\left(2\, m^2 - (\delta\omega_k)^2\right) \Tilde{J}^{(2)}_{21} + 2\, (\delta\omega_k) \Tilde{J}^{(1)}_{21}\right) 
+ \frac{1}{3} \Big(\left(3\, c_s^2 - 1 \right)  \Tilde{J}^{(1)}_{k,\, 31} + \left(m^2 - (\delta\omega_k)^2\right)  \Tilde{J}^{(3)}_{k,\, 31} \nonumber\\
&+ 2\, (\delta\omega_k) \Tilde{J}^{(2)}_{k,\, 31}\Big) - \frac{\beta\, c_s^2}{9} \left(\frac{\partial(\delta\omega_k)}{\partial\beta}\right) \! \Big( \left(3\, c_s^2 - 1\right) \Tilde{J}^{(0)}_{k,\, 10} + \left(m^2 - (\delta\omega_k)^2\right) \Tilde{J}^{(2)}_{k,\, 10} + 2 (\delta\omega_k) \Tilde{J}^{(1)}_{k,\, 10}\Big)\nonumber\\
&+ \frac{\beta\, c_s^2\, \chi_{\Pi k}}{3\, \beta_\Pi} \! \Big( \! \left(3\, c_s^2 - 1\right) \Tilde{J}^{(0)}_{k,\, 20} + \left(m^2 - (\delta\omega_k)^2\right) \Tilde{J}^{(1)}_{k,\, 10} + 2\, (\delta\omega_k) \Tilde{J}^{(0)}_{k,\, 10}\Big) \nonumber\\
&- \frac{\beta\, c_s^2\, \chi_{\Pi k}\, \chi_{3k}}{\beta_\Pi} \Tilde{J}^{(0)}_{k,\, 10}
+ \frac{\beta\, c_s^2}{3} \left(\left(3\, c_s^2 - 1\right) \Tilde{M}^{(0)}_{k,\, 30} + \left(m^2 - (\delta\omega_k)^2\right) \Tilde{M}^{(0)}_{k,\, 10} + 2 (\delta\omega_k) \Tilde{M}^{(0)}_{k,\, 20}\right) \nonumber\\
&+ \frac{\beta}{3} \! \bigg(\left(3\, c_s^2 - 1\right) \Tilde{M}^{(0)}_{k,\, 31} + \left(m^2 - (\delta\omega_k)^2\right)  \Tilde{M}^{(2)}_{k,\, 31} + 2 (\delta\omega_k) \Tilde{M}^{(1)}_{k,\, 31} \bigg)
- \beta\, \chi_{3k} \Big(c_s^2\, \Tilde{M}^{(0)}_{k,\, 20} + \Tilde{M}^{(1)}_{k,\, 31}\Big) \nonumber\\
&- \frac{2}{3} \left(\left(\frac{\partial\left(\delta \omega_{k}\right)}{\partial \beta}\right) \beta c_{s}^{2} + \left(\delta \omega_{k}\right)\right) \! \Big( \left(m^2 -(\delta\omega_k)^2\right) \Tilde{J}^{(2)}_{k,\, 10} + (\delta\omega_k) \Tilde{J}^{(1)}_{k,\, 10} \Big) \nonumber\\
&+ \frac{1}{3} \left(\left(\frac{\partial\left(\delta \omega_{k}\right)}{\partial \beta}\right) \beta c_{s}^{2} + \left(\delta \omega_{k}\right)\right) \! \Big(\left(3\, c_s^2 - 1\right) \Tilde{J}^{(0)}_{k,\, 10} + \left(m^2 - (\delta\omega_k)^2\right) \Tilde{J}^{(2)}_{k,\, 10} + 2 (\delta\omega_k) \Tilde{J}^{(1)}_{k,\, 10}\Big) \nonumber\\
&- \frac{\beta}{3} \left(\left(\frac{\partial\left(\delta \omega_{k}\right)}{\partial \beta}\right) \beta c_{s}^{2} + \left(\delta \omega_{k}\right)\right) \! \Big(\left(3\, c_s^2 - 1\right) \Tilde{M}^{(0)}_{k,\, 20} + \left(m^2 - (\delta\omega_k)^2\right) \Tilde{M}^{(1)}_{k,\, 10} + 2 (\delta\omega_k) \Tilde{M}^{(0)}_{k,\, 10}\Big) \nonumber\\
&+ \beta \left(\left(\frac{\partial\left(\delta \omega_{k}\right)}{\partial \beta}\right) \beta c_{s}^{2} + \left(\delta \omega_{k}\right)\right) \chi_{3k} \Tilde{M}^{(0)}_{k,\, 10} \Bigg]\nonumber\\
&- \frac{\beta^2}{6\, \beta_\Pi^2} \sum_k\bigg[3\, \chi_{3k}^2 \Tilde{J}^{(0)}_{k,\, 10} - 2\, \chi_{3k} \left(\!\left(3\, c_s^2 - 1\right) \Tilde{J}^{(0)}_{k,\, 20} + \left(m^2 - (\delta\omega_k)^2\right) \Tilde{J}^{(1)}_{k,\, 10} + 2 (\delta\omega_k) \Tilde{J}^{(0)}_{k,\, 10}\right) \nonumber\\
&+ \bigg(\!\!\left(3\, c_s^2 -1\right)^2 \Tilde{J}^{(0)}_{k,\, 30} + \left( m^2 - (\delta\omega_k)^2\right)^2 \Tilde{J}^{(2)}_{k,\, 10}+ 4 (\delta\omega_k)^2 \Tilde{J}^{(0)}_{k,\, 10} + 2 \left(3\, c_s^2 -1\right) \left(m^2 - (\delta\omega_k)^2\right) \Tilde{J}^{(0)}_{k,\, 10} \nonumber\\
&+ 4 \left(3\, c_s^2 -1\right) (\delta\omega_k) \Tilde{J}^{(0)}_{k,\, 20} + 4 (\delta\omega_k) \left(m^2 - (\delta\omega_k)^2\right) \Tilde{J}^{(1)}_{k,\, 10} \bigg)\bigg].\label{Lambda PiPi}
\end{align}
%
%%%%%%%%%%%%%%%%%%%%%%%%%%%%%%%%%%%%%%%%%%%%%%%%%%%%%%%%%%%%%%%%%%%%%

\bibliography{ref}{}

\end{document}